\documentclass[journal,twocolumn,10pt]{IEEEtran}
\pdfoutput=1
\usepackage{graphicx,color,epsfig,rotating}
\usepackage{float}
\usepackage{epsf}
\usepackage{latexsym}
\usepackage{caption}
\DeclareCaptionType{copyrightbox}
\usepackage{subcaption}
\usepackage{cite}
\usepackage{comment}
\usepackage{amsmath}
\usepackage{amssymb}
\usepackage{placeins}
\renewcommand{\arg}{{\rm arg}}

\begin{document}
\title{Performance Modeling of Next-Generation Wireless Networks}
\author{Antonios Michaloliakos, Ryan Rogalin, Yonglong Zhang,  Konstantinos Psounis,~\IEEEmembership{Senior Member, IEEE} and 
Giuseppe Caire,~\IEEEmembership{Fellow, IEEE}%
\IEEEcompsocitemizethanks{
\IEEEcompsocthanksitem All authors are with the University of Southern California.
}}

\date{\today}
\maketitle

%\includecomment{longver}
\excludecomment{longver}

\setlength\unitlength{1mm}

\newcommand{\insertfig}[3]{
\begin{figure}[htbp]\begin{center}\begin{picture}(120,90)
\put(0,-5){\includegraphics[width=12cm,height=9cm,clip=]{#1.eps}}\end{picture}\end{center}
\caption{#2}\label{#3}\end{figure}}

\newcommand{\insertxfig}[4]{
\begin{figure}[htbp]
\begin{center}
\leavevmode \centerline{\resizebox{#4\textwidth}{!}{\input
#1.pstex_t}}
%\vspace*{-0.2in}
\caption{#2} \label{#3}
\end{center}
\end{figure}}

\long\def\comment#1{}

% bb font symbols

\newfont{\bbb}{msbm10 scaled 700}
\newcommand{\CCC}{\mbox{\bbb C}}

\newfont{\bb}{msbm10 scaled 1100}
\newcommand{\CC}{\mbox{\bb C}}
\newcommand{\PP}{\mbox{\bb P}}
\newcommand{\RR}{\mbox{\bb R}}
\newcommand{\QQ}{\mbox{\bb Q}}
\newcommand{\ZZ}{\mbox{\bb Z}}
\newcommand{\FF}{\mbox{\bb F}}
\newcommand{\GG}{\mbox{\bb G}}
\newcommand{\EE}{\mbox{\bb E}}
\newcommand{\NN}{\mbox{\bb N}}
\newcommand{\KK}{\mbox{\bb K}}

% Vectors

\newcommand{\av}{{\bf a}}
\newcommand{\bv}{{\bf b}}
\newcommand{\cv}{{\bf c}}
\newcommand{\dv}{{\bf d}}
\newcommand{\ev}{{\bf e}}
\newcommand{\fv}{{\bf f}}
\newcommand{\gv}{{\bf g}}
\newcommand{\hv}{{\bf h}}
\newcommand{\iv}{{\bf i}}
\newcommand{\jv}{{\bf j}}
\newcommand{\kv}{{\bf k}}
\newcommand{\lv}{{\bf l}}
\newcommand{\mv}{{\bf m}}
\newcommand{\nv}{{\bf n}}
\newcommand{\ov}{{\bf o}}
\newcommand{\pv}{{\bf p}}
\newcommand{\qv}{{\bf q}}
\newcommand{\rv}{{\bf r}}
\newcommand{\sv}{{\bf s}}
\newcommand{\tv}{{\bf t}}
\newcommand{\uv}{{\bf u}}
\newcommand{\wv}{{\bf w}}
\newcommand{\vv}{{\bf v}}
\newcommand{\xv}{{\bf x}}
\newcommand{\yv}{{\bf y}}
\newcommand{\zv}{{\bf z}}
\newcommand{\zerov}{{\bf 0}}
\newcommand{\onev}{{\bf 1}}

% Matrices

\newcommand{\Am}{{\bf A}}
\newcommand{\Bm}{{\bf B}}
\newcommand{\Cm}{{\bf C}}
\newcommand{\Dm}{{\bf D}}
\newcommand{\Em}{{\bf E}}
\newcommand{\Fm}{{\bf F}}
\newcommand{\Gm}{{\bf G}}
\newcommand{\Hm}{{\bf H}}
\newcommand{\Id}{{\bf I}}
\newcommand{\Jm}{{\bf J}}
\newcommand{\Km}{{\bf K}}
\newcommand{\Lm}{{\bf L}}
\newcommand{\Mm}{{\bf M}}
\newcommand{\Nm}{{\bf N}}
\newcommand{\Om}{{\bf O}}
\newcommand{\Pm}{{\bf P}}
\newcommand{\Qm}{{\bf Q}}
\newcommand{\Rm}{{\bf R}}
\newcommand{\Sm}{{\bf S}}
\newcommand{\Tm}{{\bf T}}
\newcommand{\Um}{{\bf U}}
\newcommand{\Wm}{{\bf W}}
\newcommand{\Vm}{{\bf V}}
\newcommand{\Xm}{{\bf X}}
\newcommand{\Ym}{{\bf Y}}
\newcommand{\Zm}{{\bf Z}}

% Calligraphic

\newcommand{\Ac}{{\cal A}}
\newcommand{\Bc}{{\cal B}}
\newcommand{\Cc}{{\cal C}}
\newcommand{\Dc}{{\cal D}}
\newcommand{\Ec}{{\cal E}}
\newcommand{\Fc}{{\cal F}}
\newcommand{\Gc}{{\cal G}}
\newcommand{\Hc}{{\cal H}}
\newcommand{\Ic}{{\cal I}}
\newcommand{\Jc}{{\cal J}}
\newcommand{\Kc}{{\cal K}}
\newcommand{\Lc}{{\cal L}}
\newcommand{\Mc}{{\cal M}}
\newcommand{\Nc}{{\cal N}}
\newcommand{\Oc}{{\cal O}}
\newcommand{\Pc}{{\cal P}}
\newcommand{\Qc}{{\cal Q}}
\newcommand{\Rc}{{\cal R}}
\newcommand{\Sc}{{\cal S}}
\newcommand{\Tc}{{\cal T}}
\newcommand{\Uc}{{\cal U}}
\newcommand{\Wc}{{\cal W}}
\newcommand{\Vc}{{\cal V}}
\newcommand{\Xc}{{\cal X}}
\newcommand{\Yc}{{\cal Y}}
\newcommand{\Zc}{{\cal Z}}

% Bold greek letters

\newcommand{\alphav}{\hbox{\boldmath$\alpha$}}
\newcommand{\betav}{\hbox{\boldmath$\beta$}}
\newcommand{\gammav}{\hbox{\boldmath$\gamma$}}
\newcommand{\deltav}{\hbox{\boldmath$\delta$}}
\newcommand{\etav}{\hbox{\boldmath$\eta$}}
\newcommand{\lambdav}{\hbox{\boldmath$\lambda$}}
\newcommand{\epsilonv}{\hbox{\boldmath$\epsilon$}}
\newcommand{\nuv}{\hbox{\boldmath$\nu$}}
\newcommand{\muv}{\hbox{\boldmath$\mu$}}
\newcommand{\zetav}{\hbox{\boldmath$\zeta$}}
\newcommand{\phiv}{\hbox{\boldmath$\phi$}}
\newcommand{\psiv}{\hbox{\boldmath$\psi$}}
\newcommand{\thetav}{\hbox{\boldmath$\theta$}}
\newcommand{\tauv}{\hbox{\boldmath$\tau$}}
\newcommand{\omegav}{\hbox{\boldmath$\omega$}}
\newcommand{\xiv}{\hbox{\boldmath$\xi$}}
\newcommand{\sigmav}{\hbox{\boldmath$\sigma$}}
\newcommand{\piv}{\hbox{\boldmath$\pi$}}
\newcommand{\rhov}{\hbox{\boldmath$\rho$}}

\newcommand{\Gammam}{\hbox{\boldmath$\Gamma$}}
\newcommand{\Lambdam}{\hbox{\boldmath$\Lambda$}}
\newcommand{\Deltam}{\hbox{\boldmath$\Delta$}}
\newcommand{\Sigmam}{\hbox{\boldmath$\Sigma$}}
\newcommand{\Phim}{\hbox{\boldmath$\Phi$}}
\newcommand{\Pim}{\hbox{\boldmath$\Pi$}}
\newcommand{\Psim}{\hbox{\boldmath$\Psi$}}
\newcommand{\Thetam}{\hbox{\boldmath$\Theta$}}
\newcommand{\Omegam}{\hbox{\boldmath$\Omega$}}
\newcommand{\Xim}{\hbox{\boldmath$\Xi$}}

% Sans Serif small case

\newcommand{\asf}{{\sf a}}
\newcommand{\bsf}{{\sf b}}
\newcommand{\csf}{{\sf c}}
\newcommand{\dsf}{{\sf d}}
\newcommand{\esf}{{\sf e}}
\newcommand{\fsf}{{\sf f}}
\newcommand{\gsf}{{\sf g}}
\newcommand{\hsf}{{\sf h}}
\newcommand{\isf}{{\sf i}}
\newcommand{\jsf}{{\sf j}}
\newcommand{\ksf}{{\sf k}}
\newcommand{\lsf}{{\sf l}}
\newcommand{\msf}{{\sf m}}
\newcommand{\nsf}{{\sf n}}
\newcommand{\osf}{{\sf o}}
\newcommand{\psf}{{\sf p}}
\newcommand{\qsf}{{\sf q}}
\newcommand{\rsf}{{\sf r}}
\newcommand{\ssf}{{\sf s}}
\newcommand{\tsf}{{\sf t}}
\newcommand{\usf}{{\sf u}}
\newcommand{\wsf}{{\sf w}}
\newcommand{\vsf}{{\sf v}}
\newcommand{\xsf}{{\sf x}}
\newcommand{\ysf}{{\sf y}}
\newcommand{\zsf}{{\sf z}}

% mixed symbols

\newcommand{\sinc}{{\hbox{sinc}}}
\newcommand{\diag}{{\hbox{diag}}}
\renewcommand{\det}{{\hbox{det}}}
\newcommand{\trace}{{\hbox{tr}}}
\newcommand{\sign}{{\hbox{sign}}}
\renewcommand{\arg}{{\hbox{arg}}}
\newcommand{\var}{{\hbox{var}}}
\newcommand{\cov}{{\hbox{cov}}}
\newcommand{\SINR}{{\sf SINR}}
\newcommand{\SNR}{{\sf SNR}}
\newcommand{\Ei}{{\rm E}_{\rm i}}
\renewcommand{\Re}{{\rm Re}}
\renewcommand{\Im}{{\rm Im}}
\newcommand{\eqdef}{\stackrel{\Delta}{=}}
\newcommand{\defines}{{\,\,\stackrel{\scriptscriptstyle \bigtriangleup}{=}\,\,}}
\newcommand{\<}{\left\langle}
\renewcommand{\>}{\right\rangle}
\newcommand{\herm}{{\sf H}}
\newcommand{\trasp}{{\sf T}}
\newcommand{\transp}{{\sf T}}
\renewcommand{\vec}{{\rm vec}}
\newcommand{\Psf}{{\sf P}}
%\newcommand{\mod}{{\rm mod}}

% Colors

\newcommand{\RED}{\color[rgb]{1.00,0.10,0.10}}
\newcommand{\red}{\color[rgb]{1.00,0.10,0.10}}
\newcommand{\BLUE}{\color[rgb]{0,0,0.90}}
\newcommand{\GREEN}{\color[rgb]{0,0.80,0.20}}

%%%%%%%%%%%%%%%%%%%%%%%%%%%%%%%%%%%%%%%%%%%%%%%%
\begin{abstract}
The industry is satisfying the increasing demand for wireless bandwidth by densely deploying a large number 
of access points which are centrally managed, e.g. enterprise WiFi networks deployed in university campuses, 
companies, airports etc. This ``small cell" architecture is gaining traction in the cellular world as well, 
as witnessed by the direction in which 4G+ and 5G standardization is moving.
Prior academic work in analyzing such large-scale wireless networks either uses oversimplified models for the physical layer,
or ignores other important, real-world aspects of the problem, like MAC layer considerations, topology characteristics, and protocol overhead.
On the other hand, the industry is using for deployment purposes on-site surveys and simulation tools which do not scale, cannot 
efficiently optimize the design of such a network, and do not explain why one design choice is better 
than another.

In this paper we introduce a simple yet accurate analytical model which combines the realism and practicality 
of industrial simulation tools with the ability to scale, analyze the effect of various design parameters,
and optimize the performance of real-world deployments. The model takes into account all central system parameters,
including channelization, power allocation, user scheduling, load balancing, MAC, advanced PHY techniques 
(single and multi user MIMO as well as cooperative transmission from multiple access points), topological 
characteristics and protocol overhead.
The accuracy of the model is verified via extensive simulations and the model is used to study a wide range of
real world scenarios, providing design guidelines on the effect of various design parameters 
on performance. 

\end{abstract}
%\category{C.2}{Computer-Communication Networks}{Wireless Communication}
%\terms{Theory, Performance, Validation}
%\keywords{Analytical Model, MIMO, MU-MIMO, Distributed MIMO, CSMA} % NOT required for Procgiteedings

\section{Introduction}
\label{section:introduction}

Modern wireless devices such as tablets and smartphones are pushing the demand 
for higher and higher wireless data rates while causing significant stress to existing wireless networks. While 
successive generations of wireless standards achieve continuous improvement, it is the general understanding 
of both academic research and the industry that a significant increase in wireless traffic demand can be met only by 
a dramatically denser spectrum reuse, i.e., by deploying more base stations/access points per square kilometer,
coupled with advanced physical (PHY) layer techniques to reduce inter-cell interference.

Enterprise WiFi networks have been deployed following this paradigm for years. As a matter of fact, the density of 
access points (APs) has increased to a point where inter-cell interference is canceling any additional gains from
even denser deployments. At the same time, advanced physical layer techniques have been incorporated into the standards,
most notably single-user MIMO in 802.11n and multi-user MIMO in 802.11ac.
Cellular networks, unable to satisfy the bandwidth demand of data plans, resort to WiFi offloading, i.e. they
deploy WiFi networks to offload the cellular network.
Future cellular network architectures will most likely follow a similar pattern, that is, they will
consist of many small cells densely deployed and use advanced physical layer techniques, e.g. massive
MIMO. 
%Unfortunately, despite these efforts, the performance of such networks is still unsatisfactory, especially in crowded 
%scenarios like campus centers, conference halls, airpots, etc.

It is evident that a wireless network consisting of a large number of small, overlapping cells cannot be left to operate 
in a completely decentralized manner. The industry has responded to the need to efficiently managing such networks
with tools which are mostly based on on-site measurements, simulations, and over-simplistic analytical models.
Based on the available public information about such tools in the enterprise WiFi market \cite{cisco,aruba}, these tools perform three main operations: (i) user load balancing among APs, (ii) interference management between APs by channel allocation and 
power control, and (iii) optimization of the Clear Channel Assessment (CCA) CSMA threshold to allow for 
concurrent transmissions which can tolerate interference from nearby APs. 
While such network management tools have increased the performance of enterprise WiFi networks,
they do not scale well, cannot be used to efficiently optimize the system, and do not incorporate
the effects of advanced physical layer techniques.
	
%Academia has generated a large body of work on the analysis of such wireless networks. There are two main approaches
%mostly found in the communication theory literature. First, there is significant literature using asymptotics of random matrices to 
%extract information about achievable rates and Signal to Interference plus Noise Ratios (SINRs), see for example REF\cite{for_giuseppe} and references therein. 
%In this approach, typically users are located on a discrete and regular grid of spatial points, and base stations are on some 
%form of lattice. Second, there is a body of work which uses stochastic geometry techniques REF\cite{for_giuseppe}. In this case, the focus is on 
%the random placement of base stations and users according to Poisson point processes. The focus is on finding the tail distribution
%of the SINR for a given reference user with respect to its serving base station, in the presence of interference from other base 
%stations. Despite their sophistication, the above models cannot be used in a real-world setting for a number of reasons. They are 
%hard to use as an ``off the shelf" tool, make simplistic assumptions regarding the topology of the network, and ignore MAC layer 
%considerations and protocol overhead.

In this paper, we introduce an accurate and practical analytical model which takes into consideration 
all the important parameters affecting the performance of present and future wireless networks, and
can be efficiently used in real-world setups.
Specifically, we model and investigate the performance impact of physical layer features such as
channelization, power allocation, topological characteristics
(e.g. user density and AP distribution in various buildings/structures)
and physical layer techniques like single-user (SU), multi-user (MU), and
distributed MU-MIMO \cite{airsync_ton, rahul_jmb} (where a number of remote APs coordinate and 
transmit concurrently and jointly to multiple users).
Additionally, we model the performance impact of MAC and higher layer features such as user-AP association, 
MAC parametrization and adaptive coding/modulation.

Our main contributions are the following. We introduce and validate through simulations the first (to the best of our knowledge) analytical model 
which can be applied in real-world scenarios while taking into account all the important design parameters in the PHY, MAC and 
higher layers. Second, we apply the model to next generation wireless networking technologies such as MU-MIMO and distributed 
MU-MIMO for which there is currently no clear understanding of large scale network performance. Third, we 
apply the model to a variety of real world scenarios, including conference halls, office buildings, open spaces, large stadiums, 
etc., and study a number of important phenomena including 
the impact of different channelization options (e.g. using the whole bandwidth as a complete, 
continuous channel vs. chopping it down to multiple channels to mitigate interference), 
the effect of the CSMA CCA threshold in user throughput under the different wireless technologies, 
the performance of a well-deployed, close to ideal CSMA based network under different technologies,
the degradation of performance from using realistic quantized rates based on the Modulation and Coding schemes (MCS), etc.
(see Section \ref{section:simulations} for a detailed discussion).
	
The outline of the paper is as follows. In the next section we discuss related work. 
In Section \ref{section:primer} we discuss basic PHY characteristics.
Section \ref{section:model} motivates and describes a unified analytic treatment of wireless network deployments consisting of an analytical 
model for various current and next generation PHY layer schemes as well as for CSMA MAC. 
The validation and limitations of our model are studied via extensive simulations for tractable scenarios of interest in Section \ref{section:validation}. 
Finally Section \ref{section:simulations} applies the analytic model in various deployment scenarios of interest like conference halls, open 
and closed office floor plans and stadiums.
\section{Related Work}
\label{section:relatedwork}

The wide adoption of wireless networks both as infrastructure and especially as end-user connectivity has attracted the interest of many research communities, namely those of 
communication/information theory and
networking.

There are two main trends the communication literature follows in analyzing wireless network deployments. The first makes use of techniques based on random matrix theory to 
extract performance measures (such as achievable rates, SINRs etc.) combined with combinatorial and convex optimization methods to solve problems that appear in multi-cell 
wireless networks. Such problems include but are not limited to: finding the optimal achievable rates under power control, 
%\cite{huh-tulino-caire,lakshminaryana} 
massive MIMO system asymptotics, 
%\cite{rusek,hoydis}
base station cooperation towards a distributed MU-MIMO solution %\cite{massive_mimo_caire,gesbert,zakhour,dekerret}
and more as can be found in prior work of ours \cite{huh-tulino-caire,massive_mimo_caire,ramprashad_caire} and others 
\cite{lakshminaryana,rusek,hoydis,gesbert,zakhour,dekerret}. The second approach is based on stochastic geometry results and focus on the spatial configuration modeling of 
random placement of APs and users according some stochastic point process (see for example \cite{spatail_modelling_jindal} and references therein for such works). The developed 
techniques have been extensively applied in heterogeneous cellular networks on problems such as: load balancing \cite{andrews}, user-AP association \cite{dhillon} and K-tier base 
station modeling \cite{dhillon2}. Most of these works do not consider MIMO and advanced interference management schemes at the PHY layer since they introduce statistical 
dependence between the nodes, and this would break the independence on which most of these results are based.
% However, recent progress has been made also to incorporate more advanced PHY schemes in a stochastic geometry analysis \cite{gupta}. 
Also, a common theme of the aforementioned information theoretic literature is that they almost invariably neglect the MAC scheduling algorithms that all wireless networks make use 
of. We propose a simple analytic approach that incorporates PHY layer advances as the ones mentioned above in a single PHY/MAC layer model.

On the other hand, the networking community looked at this problem from a different angle: early work of Bianchi\cite{bianchi_csma2,bianchi_csma} on 802.11 MAC layer proposed 
an analytical model to analyze CSMA/CA overhead and performance. Meanwhile in \cite{ziouva_csma} the authors investigated the performance of exponential backoff mechanism 
in terms of throughput and delay. In \cite{csma_87,csma_boe,csma_ton,csma_infocom} similar Markov Chain models of CSMA/CA have been developed and employed to develop 
algorithms optimizing various performance metrics. Finally, in our own previous works \cite{csmaefficiency,region_ton}, a full analytic model for computing the achievable rate region 
of CSMA in multi hop wireless networks has been presented. These works are mostly based on pure upper layer modeling and do not take the advances of physical layer into 
account, which is the main contribution of this particular paper. A few recent papers have taken limited steps towards such a combination but lack the analytic simplicity our model 
has, and limit their results to standard simulation based approaches \cite{lte_vs_wifi} or neglect MAC considerations all together \cite{goldsmith}.

Lastly, there are a number of tools that the industry currently uses for wireless network deployment guidance. For example, Fluke Networks has developed a product \cite{fluke} which creates a model for the wireless environment so that an administrator may simulate and predict the performance. These tools mostly resort to real time 
on-site surveys which do not scale and do not incorporate state-of-the-art advances in the PHY layer.

%With the exponentially increasing number of wireless devices, there has been enormous amount of work to evaluate 802.11 wireless systems' performance based on physical layer. 
%To be more precisely, there are generally 
%two different approaches: the first approach uses techniques based on asymptotic of random matrices with combinatorial optimization and convex optimization in order to solve 
%problems in typical multi-cell wireless networks.

%For example, finding the optimal achievable rates under power control \cite{zfb_caire,lakshminaryana,goldsmith}, MIMO system scaling\cite{rusek,hoydis}, base station corporation 
%\cite{massive_mimo_caire,gesbert,zakhour,dekerret}, optimal spatial configuration\cite{spatail_modelling_jindal} and more. 

%The second approach is based on stochastic geometry and is focusing on maximizing SINR of a given reference user under a heterogenous wireless network with random base 
%station placement and Poisson user access, 
%which resolved problems such as load balancing\cite{gupta,andrews}, user association\cite{dhillon} and K-tier base station modeling\cite{dhillon2}. As we could see, problems in 
%multi-cell and heterogeneous wireless networks 
%with MIMO, MU-MIMO
%and distributed MU-MIMO has been deeply studied in the above works from the point of view of communication and information theories. However, their analytical results are often 
%complicated and not easy to be implemented 
%in practice.
\section{Next-Generation Wi-Fi Primer}
\label{section:primer}

\label{sec:phyprimer}
{\bf OFDM primer.}
Orthogonal Frequency Division Multiplexing (OFDM) \cite{molisch-book} is the preferred PHY layer modulation 
technique of modern cellular and WLAN networks. 
OFDM consists of taking blocks of $N$ coded modulation ``frequency domain'' symbols 
and transforming them into a block of $N$ ``time domain chips''
via an Inverse Discrete Fourier Transform (IDFT). Each block of $N$ time-domain chips is expanded by repeating the last $L$ chips at the beginning of the block. 
This precoding technique, known as Cyclic Prefixing (CP), is able to turn the linear convolution of the transmitted signal with the channel impulse response
into a block-by-block cyclic convolution, provided that the length of the multipath channel (in chips) is not larger than $L$. 
At the receiver, after block timing and carrier frequency synchronization, the CP is removed and a DFT is applied 
to the resulting blocks of $N$ chips in order to recover the $N$ frequency domain symbols. 
As a result, the time-domain multipath channel is transformed into a set of parallel frequency-flat 
channels (referred to as subcarriers) in which each frequency-domain  symbol experiences only a complex multiplicative ``fading'' channel 
coefficient, corresponding approximately to the channel transfer function evaluated at each subcarrier 
center frequency (see \cite{ryan-twc} for a recent accurate and general 
model of OFDM including also non-ideal transmit/receive effects).  

{\bf Topology and pathloss model.} 
We define a region based on some topology and a fine, regular grid of quantized coordinates, 
with some geometry of walls and blocking objects (e.g., buildings in an outdoor campus scenario, or office walls in an indoor building scenario).  
Access points (APs) and user terminals (UTs) are located at the nodes of the grid, and are placed in scenario-dependent locations.
\begin{longver}
{\red The pathlosses between APs and UTs are determined using the WINNER-II model \cite{winner2}, 
which takes distance and log-normal shadowing into account for different scenarios (indoor, outdoor, small rooms, large halls, etc.).}
\end{longver}

{\bf Small-scale fading.}
It is well-known that typical wireless channels in WLAN environments are slowly-varying in time. 
For a sequence of successive time slots, each one corresponding to a data packet, 
the channel coefficients are strongly correlated such that estimates obtained at a given point in time are accurate 
for a fairly large number of time slots \cite{molisch-book}. 
\begin{longver}
{\red For example, for a typical nomadic WLAN user moving at speed $v$ less 
than 1 m/s, at $f_0 = 2.4$ GHz carrier frequency, the so-called Doppler bandwidth  is at most $B_d = f_0 v/c = 8$ Hz ($c$ denoting the speed of light). 
Then, the time over which the channel remains approximately constant (coherence time)
is given by at least $T_c \approx 1/B_d = 125$ ms. }
For example, considering that a data packet in the 802.11 has duration at most 5 ms and the coherence time of a typical WLAN channel is in the order of 125 ms \cite{molisch-book}, the fading channel remains approximately constant over blocks of at least 25 slots. 
\end{longver}

\begin{longver}
{\red As said before, with OFDM each system channel can be represented as a set of $N$ subcarriers. 
For channels of 20 MHz of bandwidth, with $N = 64$ and CP length $L = 16$, the duration of an OFDM symbol
is $N + L = 80$ chips $\times 0.05 \times 10^{-6}$ s/chip $= 4 \mu s$ (notice: the duration of each chip is equal to the reciprocal of the channel bandwidth). 
Hence, the fading channel remains approximately constant over $125/4 \times 10^3 = 31250$ OFDM symbols. It follows that 
by devoting a small fraction of such symbols to pilots for channel estimation it is possible to accurately estimate the channel coefficients
and enable coherent detection and channel state information feedback at the cost of a small overhead \cite{molisch-book}. } 
\end{longver}

Motivated by this, in the following we assume that the frequency-domain channel 
coefficients are random but constant over many data slots, and 
drop the time index when referring to a time slot for the sake of notation simplicity. 
We shall use index 
$\nu$ to indicate the OFDM subcarrier index.
Hence, the channel between an AP $i$ equipped with $M$ antennas and a user terminal (UT) $k$ equipped 
with a single antenna at any given generic time slot and OFDM subcarrier $\nu$ is given by 
\begin{equation} \label{dl-channel} 
y_k[\nu] = \sqrt{g_{ik}} \xv_i[\nu]^\herm \hv_{ik}[\nu] + z_k[\nu], 
\end{equation}
where $\hv_{ik}[\nu]$ is an $M \times 1$ vector with Gaussian i.i.d. elements 
$\sim \Cc\Nc(0,1)$,\footnote{$h \sim \Cc\Nc(0,1)$ indicates a complex circularly symmetric Gaussian
random variable with mean 0 and variance $\EE[|h|^2] = 1$. This type of small-scale fading is usually referred to as Rayleigh fading \cite{molisch-book}.} 
representing the channel coefficients between the antenna array of AP $i$ and the antenna of UT $k$,  
$\xv_i[\nu]$ is a $M \times 1$ vector containing the (frequency-domain) coded modulation symbols 
sent by AP $i$ in the downlink, on subcarrier $\nu$,  and $z_k[\nu]$ collects noise plus the interference caused by other APs
transmitting on the same time slot and frequency channel at the receiver of UT $k$. 
The channel gain $g_{ik}$ is a function of the distance between AP $i$ and UT $k$ and of other ``large-scale'' effects, such as blocking objects, walls and trees. 
This ``large-scale'' channel gain is frequency-flat, i.e., it does not depend on  the subcarrier index $\nu$, and it is modeled using
the widely accepted WINNER-II model \cite{winner2}. 

{\bf MIMO primer.}
Point-to-Point MIMO \cite{foschini-gans, telatar} consists of the techniques of extracting higher capacities from a given wireless bandwidth 
by adding antennas to the transmitter and receiver.  MIMO is ideally suited to work together with OFDM. 
We describe the effective channel between a transmitter with $N_t$ antennas and a receiver with $N_r$ antennas 
via a set of $N_t \times N_r$ channel matrices $\Hm_{ik}[\nu]$ for all the subcarriers forming the given system channel. 
One of the advantages of having multiple transmit/receive antennas is the ability to increase the effective SNR, known as a power gain. 
For example, using maximal ratio combining \cite{molisch-book} in a $1 \times N_r$ channel, the effective received signal SNR is multiplied by $\|\hv_{ik}[\nu]\|^2$. 
In the downlink, if AP $i$ has $N_t > 1$ antennas and UT $k$ has only a single antenna, knowledge of the $N_t \times 1$ downlink channel 
vector $\hv_{ik}[\nu]$ can be used to perform conjugate beamforming (aka maximal ratio beamforming) 
in order to achieve the same  SNR gain $\|\hv_{ik}[\nu]\|^2$ \cite{molisch-book}.  
In addition, a variety of techniques globally known as  {\em spatial precoding} \cite{molisch-book} enable multiplexing in the spatial domain, 
such that the capacity at high SNR is increased by  a {\em multiplicative factor} up to $\min(N_t,N_r)$, referred to as ``multiplexing gain'' 
or ``degrees of freedom (DoFs) gain.''

With Point-to-Point MIMO, a multiplexing gain larger than 1 is achievable only if multiple antennas are used on both the transmitter and the receiver side (i.e., 
$\min\{N_t, N_r\} > 1$). Particularly for the downlink this may represent a significant limitation since, as mentioned before, $N_r$ may be small (1 or at most 2). 
In this case, a set of techniques known as multiuser MIMO (MU-MIMO) \cite{caire-shamai} can be used to attain higher multiplexing gain up to
$\min\{M, KN_r\}$, for a system where an AP with $M$ antennas serves $K$ UTs with $N_r$ antennas each.  
Since the UTs do not cooperate, most MU-MIMO schemes require channel state information at the transmitter (CSIT) in order 
to precode the data streams. This allows multiple data streams to be simultaneously transmitted on the same time-frequency slot, and yet interfere in a benign and controlled manner 
at each desired UT. For instance, when using zero-forcing beamforming (ZFBF) \cite{caire-kobayashi-jindal},
knowledge of the channel matrix can be exploited in order to precode the data such that each UT receives 
only its intended data symbol, and all other interfering streams 
form the same AP cancel out (see Section \ref{section:mumimo} for more details). 

\section {Modeling of Next Generation Wireless}
\label{section:model}

\subsection{Common System Level Parameters}
\label{sec:common_params}

\begin{longver}
\begin{figure}[htbp]
   \centering
   \includegraphics[width=0.4\textwidth]{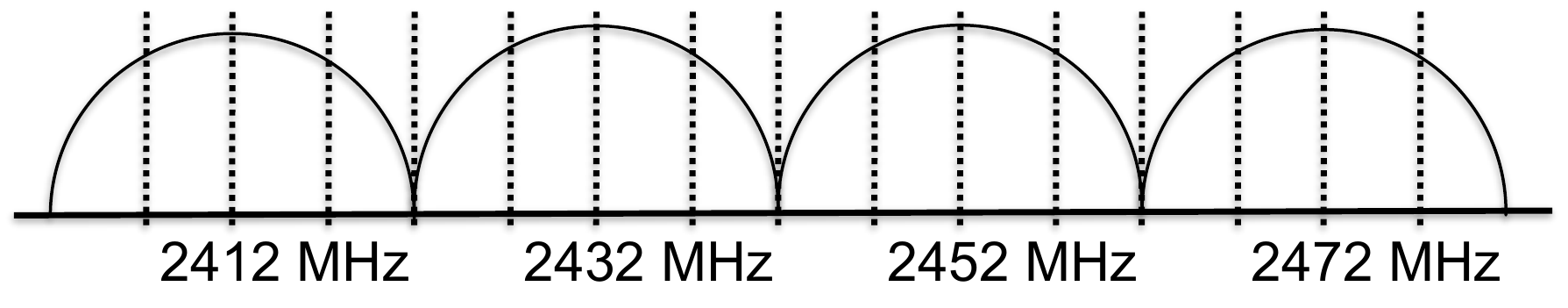} % requires the graphicx package
   \caption{Channelization for the 2.4 GHz band as provisioned by the 802.11 standard \cite{802.11ac}, with 4 non-overlapping channels
   (namely, channel 1, 5, 9 and 13) each of which has bandwidth 20MHz.}
   \label{fig:channels}
\end{figure}
\end{longver}
{\bf Frequency Reuse/Channel Assignment.}
In our numerical examples,  we have assumed that the system has 80 MHz of total bandwidth that can be partitioned into 20 MHz or 40 MHz 
non-overlapping channels or a single 80 MHz channel, with reference 
to the 802.11 2.4GHz band and the 802.11ac revision \cite{802.11ac}. 
Each AP is active on one of the system channels. 
\begin{longver}
{\red We have considered the cases where each AP uses one out of 4 non-overlapping 20 MHz channels, 
each AP uses one out of 2 non-overlapping 40 MHz channels, or all APs use the single 80 MHz channel.  In the first two cases,} 
\end{longver}
The same orthogonal channel can be allocated to multiple APs according to some suitable spatial frequency reuse scheme. 
The allocation of channels to APs is, in general, a hard problem. 
In this work, we have adopted a one-pass, greedy algorithm. 
APs are ordered according to a random sorting permutation $\pi$ and choose their channel sequentially, 
such that AP $\pi(i)$ makes its choice at step $i$ and chooses channel 
$c_{\pi(i)} \in \Cc$ where $ \Cc$ is the set of available (non-overlapping) system channels (numbered as $1\dots |\Cc|$). 
The channel choice is made according to the rule:
$$ c_{\pi(i)} = \arg \min _{c \in \Cc} \left \{ \sum_{j < i : \pi(j) \in {\Ac_{c}(i)}} I_{\pi(i),\pi(j)}^c \right \} $$
where $I_{\ell,r}^c$ denotes the interference power received by AP $\ell$ from AP $r$ in channel $c$, 
and $\Ac_{c}(i)$ denotes the set of APs already assigned to channel $c$ at step $i$ of the assignment procedure. 

{\bf User-AP Association.}
The simplest, most widely used user-AP association algorithm connects each user to the AP with the highest Received Signal Strength Indication (RSSI).
This approach does not take into account load balancing since in the presence of a non-uniform user spatial distribution, some APs
may be cluttered by a lot of users while other APs may be left almost idle.
The typical industry solution to this problem is to ``force'' a new user to switch to another AP if the AP with the strongest RSSI has already too many users, and support client roaming between APs  when conditions change \cite{cisco}. 

The AP association and load balancing issue are relatively well-studied problems in academia, see, for example, \cite{fairness,andrews,dhillon2}.
In our examples (see Section \ref{section:simulations}), we have considered a heuristic user-AP assignment scheme aimed at improving the system fairness.
Inspired by the work in \cite{userap_nsdi}, we allocate users based on  ``available capacity.'' 
A central controller orders the users according to a sorting random permutation $\pi'$ and associate each user sequentially to the AP that offers the largest 
available capacity.  In particular, the association algorithm starts by assigning to all APs  $i = 1, \ldots, N_a$ 
an empty set $\Sc_i = \emptyset$ of associated users. Then, at step $k$, user $\pi'(k)$ is added to the set $\Sc_{i_k}$ if $i_k$ is the AP index satisfying
\[ i_k = \arg \max_{i = 1, \ldots, N_a} \frac{ C_{i,\pi'(k)}}{|\Sc_i \cup \{\pi'(k)\}|}, \]
where $C_{i,k}$ denotes the peak rate of the link from AP $i$ to user $k$. 
The association ends when all users have been assigned. Notice that we use the ratio $\frac{C_{i,\pi'(k)}}{|\Sc_i \cup \{\pi'(k)\}|}$ as a proxy 
of the  {\em average} throughput that user $\pi'(k)$ can get from AP $i$. This is motivated by the fact that, under proportional fairness scheduling, 
each AP allocates its downlink channel resource in equal proportion to its associated users (equal air time policy). 

Note that while the channel allocation and user-AP association schemes adopted 
in this paper are sensible approaches, they do not correspond necessarily to optimal strategies. Nevertheless, 
the purpose of this paper is not that of proposing system optimization strategies, rather, our focus is to develop a general analytical model
which works irrespectively of the specific channel allocation and user-AP association scheme and can be used to enable such type of system optimization.

\subsection{CSMA Modeling}
\label{sec:macprimer}
%{\bf CSMA modeling.} 
In order to model CSMA, we focus on the downlink and assume that only APs compete with each other for the wireless medium. 
The APs form a contention graph by setting an SINR threshold below which the AP-to-AP interference is treated as noise. (This is known as
the Clear Channel Assessment (CCA) threshold.) 
APs in the sensing range of each other that operate on the same channel cannot operate simultaneously, due to the CSMA/CA algorithm and thus 
they share a link in the contention graph. 
%Notice that CSMA does not affect a fully cooperative distributed MU-MIMO system (see section \ref{dmumimo}).

We start with an idealized 802.11 CSMA model as in \cite{csma_87,csma_ton,csma_boe,csma_infocom}. 
Since our goal is to deduce an analytically tractable yet reasonably accurate model (see Section \ref{section:validation} for a validation of the model
against NS-2 simulations \cite{ns2}), we assume that 
%there are no hidden node effects in our network, 
the transmission and countdown times are exponentially distributed with means $1/\mu$ and $1/\lambda$ respectively, and the medium 
can be sensed instantaneously, thus collisions between APs can be avoided. 
%The first assumption is realistic when deployments seek to avoid hidden terminal setups, 
%which is the case in hidden-node-free designed networks \cite{hnremoval}. 
The first assumption makes the system easy to analyze 
since it can be modeled as a Continuous Time Markov Chain (CTMC) and can be relaxed to milder and more natural system assumptions 
without significantly affecting the results, as shown in \cite{csma_boe}. 
The second assumption might lead to an overestimation of the throughput of CSMA when the system spends a sizable time in collisions. 
However, in practical and well-designed topologies, as in optimized enterprise WiFi networks, this is not the case. 
It is important to note that our CSMA model operates on top of an $\SINR$-based PHY model and thus we fully model
phenomena like hidden terminals, channel capture, etc.

The states of the aforementioned CTMC are all the different feasible transmission patterns for the $N_a$ APs (that is, only non-conflicting APs, or, equivalently, independent sets of APs, can transmit at the same time) 
and the transitions between the different states happen when an AP that is not conflicting with the APs currently transmitting, gets out of countdown 
and starts transmitting, or when an AP finishes its transmission.
Let $\mv=\{m_k | k=1,\dots,N_a\}$ be the binary state vector of the CTMC,
where we let $m_i = 1$ if AP $i$ is transmitting and $m_i = 0$ otherwise. Let $\Mc \subseteq\{0,1\}^{N_a}$ be the state space 
of the CTMC (independent sets of the interference conflict graph).
It can be shown that the limiting stationary distribution of state $\mv \in \Mc$ is given by:
\begin{equation} \label{ctmc}
\pi_{\mv}=\frac{\rho^{\|\mv\|_1}}{\sum_{\mv' \in \Mc} \rho^{\|\mv'\|_1}},
\end{equation}
where $\rho=\lambda/\mu$ and $\|\cdot\|_1$ is the $\bf L_1$ norm, 
thus $\|\mv\|_1$ is the number of APs transmitting during state $\mv$.

\begin{figure}
%\begin{minipage}{.4\textwidth}
  \begin{subfigure}[b]{.2\textwidth}
   \centering
 %  \rule{6.4cm}{3.6cm}
     \includegraphics[scale=0.4]{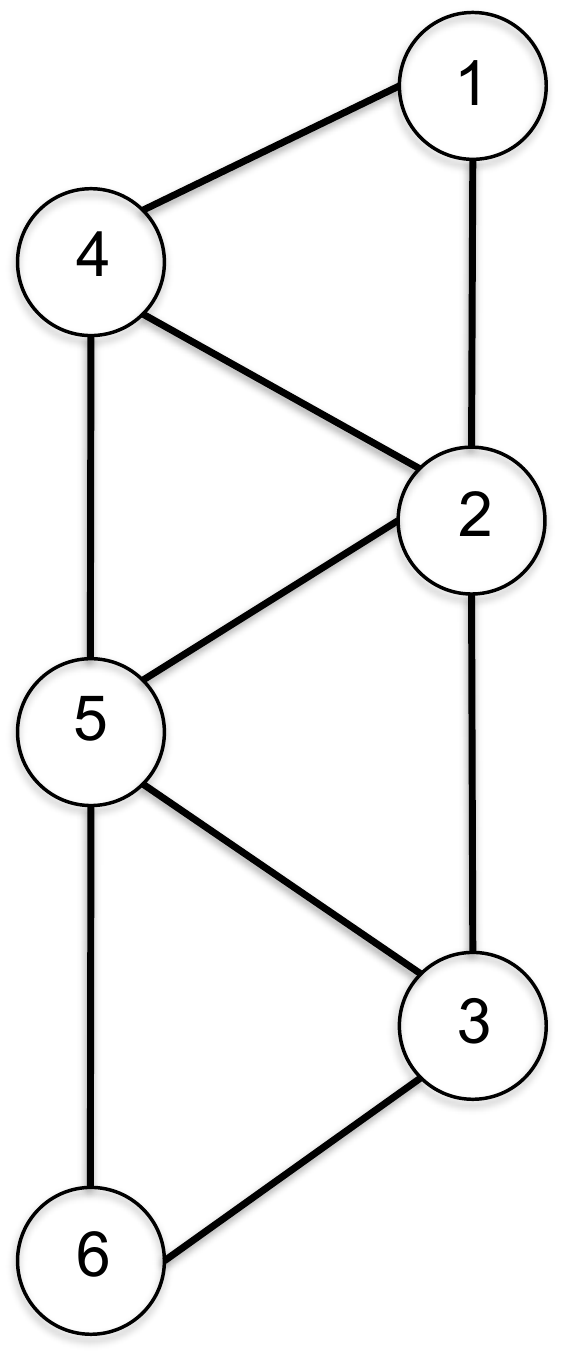} 
  \captionof{figure}{Interference graph}
  \label{fig:interf_graph}
\end{subfigure}
 \begin{subtable}[b]{.2\textwidth}
\centering
   \begin{tabular}{| c | c |  }
    \hline
%    State & 000000 & 100000 & 010000 &  $\vdots$ & 000001 &  101000 & 100010 & 100001 & 010001 &001100 &  000101 \\ \hline
%    Limiting Probability & $1/(1+6\rho+6\rho^2)$ & $\rho/(1+6\rho+6\rho^2)$ & $\rho/(1+6\rho+6\rho^2)$ & $\vdots$ & $\rho/(1+6\rho+6\rho^2)$ & $\rho^2/(1+6\rho+6\rho^2)$ & $\rho^2/(1+6\rho+6\rho^2)$ & $\rho^2/(1+6\rho+6\rho^2)$ & $\rho^2/(1+6\rho+6\rho^2)$ & $\rho^2/(1+6\rho+6\rho^2)$ & $\rho^2/(1+6\rho+6\rho^2)$  \\  \hline
    State & Limiting Probability \\ \hline
    000000 & $1/(1+6\rho+6\rho^2)$   \\ \hline
    100000 & $\rho/(1+6\rho+6\rho^2)$ \\ \hline
    010000 & $\rho/(1+6\rho+6\rho^2)$ \\ \hline
    $\vdots$ & $\vdots$  \\ \hline
    000001 & $\rho/(1+6\rho+6\rho^2)$ \\ \hline
    101000 & $\rho^2/(1+6\rho+6\rho^2)$ \\ \hline
    100010 & $\rho^2/(1+6\rho+6\rho^2)$ \\ \hline
    100001 & $\rho^2/(1+6\rho+6\rho^2)$  \\  \hline
    010001 & $\rho^2/(1+6\rho+6\rho^2)$  \\  \hline
    001100 & $\rho^2/(1+6\rho+6\rho^2)$  \\  \hline
    000101 & $\rho^2/(1+6\rho+6\rho^2)$  \\  \hline
    \end{tabular}
    
%  \end{center}
  \caption{Limiting probabilities of the CSMA CTMC}
  \label{tbl:stationary}
\end{subtable}
\caption{AP interference graph and limiting probabilities of the corresponding CSMA CTMC.}
%\end{minipage}
\end{figure}

As an example, in Figure \ref{fig:interf_graph} we see 6 APs operating on the same channel and the corresponding contention graph. 
There are 13 feasible states for this graph and the probabilities for each state can be seen in Figure \ref{tbl:stationary}. 
Intuitively, one expects that larger independent sets are more likely to be scheduled by CSMA, since once an AP is selected, all sets
not containing this AP will not be scheduled. Indeed, the probabilities in Figure \ref{fig:interf_graph} are a lot larger for larger independent sets
because the transmission times are a lot larger than the countdown times and thus $\rho \gg 1$.
As a result, in large networks it is reasonable to take into 
account only states corresponding to {\em maximal independent sets} of the interference graph, rather than all feasible states. It should be noted here that identifying the maximal independent sets of a graph is known to be an NP-complete problem, nevertheless given the specific characteristics of each topology (AP density, symmetric placement etc) we have successfully run large examples in a few minutes. 
%In fact, as noted in \cite{csma_boe} it is doubtful that we can avoid this complexity without losing part of our model's validity.

Having computed the fraction of time that the network spends in each transmission pattern $\mv \in \Mc$, we next take 
into account the throughput resulting from CSMA/CA. To do so, we evaluate the total 
spectral efficiency (or each individual UT spectral efficiency) achieved for each transmission pattern as shown in the rest of the section, 
average these values with respect to the stationary distribution (see Equation \ref{ctmc}), and multiply the result by the 
channel bandwidth to obtain the throughput in bit/s.  
%In this way, we can also evaluate the impact of  using CSMA/CA with respect to ideal resource allocation when 
%the overhead due to collisions is negligible.

%The above can be simplified even further when $\rho>>1$, that is when the transmission time is significantly larger than the countdown time. When that is assumed, most time is
%spent in the maximum independent set (MIS)  states of the contention graph (the last 6 states in Table \ref{tbl:stationary} and thus we can identify the long-run proportion of time
%an AP has the channel through the following procedure\cite{csma_boe} :
%
%\begin{enumerate}
%\item Find all the maximum independent sets (MIS) for the contention graph.
%\item The long-run proportion of time the CTMC spend in state $i$ by the ratio $1/n$ where $n$ is the total number of MIS when $i$ is an MIS and $0$ otherwise.
%\end{enumerate}

\subsection{Single-user Beamforming}

%[NOTE: WE SHOULD NOT CALL THIS ``SU-MIMO'', SINCE IN THE PRIMER WE HAVE TALKED ABOUT POINT TO POINT (SINGLE USER) MIMO
%INCLUDING THE CASE OF $\min\{N_t,N_r\} > 1$. HERE, SPECIFICALLY, WE TALK ABOUT SINGLE USER BEAMFORMING. 
%MAYBE FOR THE BENEFIT OF THE READER YOU MAY WISH TO SAY (OR IN THE RESULTS SECTION) THAT SINGLE-USER BEAMFORMING IS
%MORE OR LESS WHAT IS DONE BY 11n, IN MOST VENDOR IMPLEMENTATIONS]

{\bf User rates with single-user beamforming.} 
Without loss of generality, let AP $i$ be active on channel $c_i$, and let $\nu = 1, \ldots, N$ denotes the 
corresponding OFDM subcarriers. 
Assume that UT $k$ is associated with AP $i$, i.e., $k \in \Sc_i$. 
The {\em instantaneous peak rate} at which AP $i$ can serve user $k$ is given by 
\begin{equation} \label{inst-user-rate} 
C_{ik}^\mv =  m_i \frac{1}{N} \sum_{\nu=1}^N \log \left ( 1 + \SINR_{ik}^\mv[\nu] \right ), 
\end{equation}
where $\mv$ denote the current state of the CMTC (such that this rate is zero if $m_i = 0$, i.e., AP $i$ is not
in the current active set according to the CSMA protocol). The rate $C_{ik}^\mv$ is measured in bit/s/Hz, and it is referred to as ``instantaneous'' 
since it is a function of the realization of the fading channel coefficients. The term $\SINR_{ik}^\mv[\nu]$ denotes the Signal to Interference plus Noise Ratio
on subcarrier $\nu$. With single-user beamforming, this is given by 
\begin{equation} \label{SINR} 
\SINR_{ik}^{\mv}[\nu] = \frac{g_{ik} \| \hv_{ik}[\nu] \|^2 P_i}{1 + \sum_{j \in \Ac_{c_i} : j \neq i} m_j g_{jk} \left | \vv_j^\herm[\nu] \hv_{jk}[\nu] \right |^2 P_j},
\end{equation}
where, as defined before, the set $\Ac_{c_i}$ includes the APs assigned to channel $c_i$, 
$P_i$ denotes transmit power spectral density (energy per frequency domain symbol) of AP $i$ and $\vv_i[\nu] = \hv_{ik}[\nu]/\|\hv_{ik}[\nu]\|$ is 
the unit-norm transmit beamforming vector of AP $i$ serving user $k$, 
according to the conjugate beamforming scheme (see Section \ref{sec:phyprimer}). The interfering AP $j \in \Ac_{c_i}$ uses 
the same scheme, with a unit vector $\vv_j[\nu]$ that depends on the channel vector to its own intended user. 
In the SINR denominator, the inner product $\vv_j^\herm[\nu] \hv_{jk}[\nu]$ corresponds to the ``spatial coupling'' between such vector and the channel vector
from AP $j$ and user $k$. It is important to notice that $\vv_j[\nu]$ and $\hv_{jk}[\nu]$ are statistically independent. 
Instead, the coupling between the beamforming vector $\vv_i[\nu]$ of AP $i$ and its intended user channel $\hv_{ik}[\nu]$ yields
$\vv_i^\herm[\nu] \hv_{ik}[\nu] = \|\hv_{ik}[\nu]\|$ which yields the beamforming gain at the numerator of the SINR. 
It follows that $\left | \vv_j^\herm[\nu] \hv_{jk}[\nu] \right |^2$ is a chi-squared
random variable with 2 degrees of freedom, while $\| \hv_{ik}[\nu] \|^2$ is a chi-squared random variable 
with $2M$ degrees of  freedom. 

In order to obtain simple deterministic analytical formulas for the rates,
we consider that $M$ is large enough such that the effect of small-scale fading disappears. 
Hence, $\|\hv_{ik}[\nu]\|^2 \rightarrow M$  (beamforming gain equal to the number of antennas per AP), 
and $\left | \vv_j^\herm[\nu] \hv_{jk} [\nu] \right |^2 \rightarrow 1$. 
The resulting approximated deterministic rate formula for the peak rate is:
\[ C_{ik}^\mv = m_i \log \left ( 1 + \frac{g_{ik} M P_i}{1 + \sum_{j \in \Ac_{c_i} : j \neq i} m_j g_{jk} P_j} \right ). \]
With orthogonal downlink multiple access (TDMA)  the rate region for the downlink of AP $i$ for $m_i = 1$ 
(i.e., in the time slots where AP $i$ is allowed to transmit because of CSMA/CA) is given by  the set of non-negative 
rates $\{R_{ik}^\mv : k \in \Sc_i\}$ such that
\begin{equation} \label{TDMAregion} 
\sum_{k \in \Sc_i} \frac{R_{ik}^\mv}{C_{ik}^\mv} \leq 1. 
\end{equation}
With proportional fairness downlink resource allocation, the system operates at the rate point
\begin{equation}  \label{rate-su-mimo} 
 R_{ik}^\mv = \frac{m_i}{|\Sc_i|} \log \left ( 1 + \frac{g_{ik} M P_i}{1 + \sum_{j \in \Ac_{c_i} : j \neq i} g_{jk} P_j} \right ).
 \end{equation}
Eventually, the average spectral efficiency of user $k \in \Sc_i$ when also averaging with respect to the 
stationary distribution of the CTMC that describes the CSMA/CA MAC layer, is given by 
\begin{equation} \label{MC-average} 
R_{ik} = \sum_{\mv \in \Mc} \pi_\mv  \cdot R_{ik}^\mv. 
\end{equation}
In order to convert this number in the more usual average throughput in bit/s, 
it is sufficient to multiply $R_{ik}$ by the channel bandwidth $W_{c_i}$ (measured in Hz) of the channel $c_i$ 
allocated to AP $i$.

%{\bf Throughput cumulative distribution function:} 
Once the user spectral efficiencies $R_{ik}$ are determined, we can present the results in terms of the throughput
CDF for a given placement of the APs and of the UTs, pathloss realization,  channel allocation and user-AP association. 
Letting $T_{ik} = W_{c_i}  R_{ik}$ denote user $k$ average throughput in bit/s, 
the throughput CDF over the user population is given by 
\begin{equation} \label{eqn:cdf}
F_T(r) = \frac{1}{\sum_i |\Sc_i|} \sum_{i,k} 1 \left \{ T_{ik} \leq r \right \}.
\end{equation}
It follows that $F_T(r)$ indicates the fraction of users with throughput less or equal 
to some number $r$. 
\begin{longver}
{\red From the above CDF we can extract how many users  have throughput below some threshold 
(throughput outage probability), or above some other threshold (fraction of users able to stream video, for example), 
 we can extract the average throughput, the median throughput and so on.  }
\end{longver}

%%%%%%%%%%%%%%%%%%%%%%%%%%%%%%%%%%%%%%%%%%%%%%%%
\subsection{Concentrated MU-MIMO}
\label{section:mumimo}

We now consider the case where each AP $i$ implements 
MU-MIMO in order to serve its associated users $k \in \Sc_i$. 
This scheme is inspired by the MU-MIMO mode of 802.11ac \cite{802.11ac}. 
We denote this scheme as ``concentrated'' MU-MIMO in order to stress the difference with respect to
a distributed MU-MIMO approach that shall be treated in the next section, 
and may be regarded as a future trend of WLANs and small cell networks.  
The main difficulty here is represented by the fact that AP $i$ can simultaneously serve 
a subset of users of size not larger than $\min\{M, |\Sc_i|\}$. 
Consistent with the 802.11ac  standard \cite{802.11ac}, we consider MU-MIMO based on ZFBF.

Let $\widehat{S}_i \subseteq \Sc_i$ denote the subset of users to be served on a given time slot in MU-MIMO mode, and let
$S_i = |\widehat{\Sc}_i|$ indicate its size. The $M \times 1$ channels of users $k \in \widehat{\Sc}_i$ are assumed to be known
at the AP $i$ transmitter through some form of channel state feedback (e.g., as specified in the 802.11ac standard). 
Such channel vectors are collected as the columns of a $M \times S_i$ channel matrix 
$\Hm_i[\nu] = [\hv_{i1}[\nu], \ldots, \hv_{ik}[\nu]]$, for all the subcarriers $\nu$ forming channel $c_i$, as defined before. 
The ZFBF precoded signal vector on subcarrier $\nu$ is given by 
$\xv_i[\nu] = \Vm_i[\nu] \uv_i[\nu]$, where $\uv_i[\nu]$ is a $S_i \times 1$ columns vector of frequency-domain coded modulation symbols
to be sent to users $k in \widehat{\Sc}_i$ and $\Vm_i[\nu]$ is the ZFBF precoding matrix, of dimension $M \times S_i$. 
Following \cite{caire-kobayashi-jindal}, this is given by 
\begin{equation} 
\Vm_i[\nu] = \Hm_i[\nu] \left ( \Hm^\herm_i[\nu] \Hm_i[\nu] \right )^{-1} \Xim^{1/2}_i[\nu]
\end{equation}
where $\Xim_i[\nu]$ is a column-normalizing diagonal matrix included in order to preserve the total AP transmit power.
In particular, the $k$-th diagonal element of $\Xim_i[\nu]$ is given by 
\[ \xi_{ik}[\nu] = \frac{1}{\left [ \left ( \Hm^\herm_i[\nu] \Hm_i[\nu] \right )^{-1} \right ]_{kk}} \]
where $[\cdot]_{kk}$ denotes the $k$-th diagonal element of the matrix argument.
Using the fact that $\Vm_i^\herm[\nu] \Hm_i[\nu] =  \Xim^{1/2}_i[\nu]$ and letting $\lambda_{ik}[\nu] = g_{ik} \xi_{ik}[\nu]$ denote the
effective channel coefficient including also the effect of the large-scale gain, 
from Equation (\ref{dl-channel}), the signal received at UT $k$ takes on the form
\begin{eqnarray} \label{dl-zfbf} 
y_k[\nu] & = & \sqrt{g_{ik}} \uv_i[\nu]^\herm \Vm_i^\herm[\nu] \hv_{ik}[\nu] + z_k[\nu] \nonumber \\
& = & \sqrt{\lambda_{ik}[\nu]} u_{ik}[\nu] + z_k[\nu],
\end{eqnarray}
where it is apparent that the multi-access interference from signals generated by AP $i$ and sent to the other users $k' \in \widehat{\Sc}_i$ with $k' \neq k$ 
is completely eliminated by ZFBF precoding.  

{\bf User rates with ZFBF.} 
Based on the effective channel (\ref{dl-zfbf}), we have that the instantaneous user rate takes on 
the same form (\ref{inst-user-rate}) with a different expression of $\SINR_{ik}^\mv[\nu]$. In particular, in the case of concentrated MU-MIMO
with ZFBF precoding this is given by 
\begin{equation} \label{SINRmumimo} 
\SINR_{ik}^{\mv}[\nu] = \frac{ \lambda_{ik}[\nu]  P_i/S_i}{1 + \sum_{j \in \Ac_{c_i} : j \neq i} m_j g_{jk} \| \Vm^\herm_j[\nu] \hv_{jk}[\nu]\|^2  P_j/S_j},
\end{equation}
where we have assumed (for the sake of simplicity) uniform power allocation over the downlink data streams, i.e., 
each user gets $1/S_i$ of the total AP transmit power.

A major difficulty here is represented by the fact that the coefficient  $\lambda_{ik}[\nu]$ 
depends on the channel matrix through the whole selected group of users $\widehat{\Sc}_i$, which means that the individual user 
rates do not ``decouple'': for finite dimension,  we need to calculate the rates for all the 
$G_i = {|\Sc_i| \choose S_i }$ possible 
groups.  In order to alleviate this problem and obtain a ``decoupled'' rate expression, 
we resort to asymptotic formulas.
Inspired by prior work of ours \cite{huh-tulino-caire}, %(also see \cite{Hochwald02space})
we consider the regime where both $M$ and $S_i$ become large while keeping the ratio $S_i/M \leq 1$ and fixed.
Specifically, we scale all channel coefficients by 
$1/\sqrt{M}$ and multiply the transmit power by a factor $M$, which yields
\[ \lambda_{ik}[\nu] \rightarrow \left (1 - \frac{S_i-1}{M} \right ) g_{ik}, \]
and
\[ \| \Vm^\herm_j[\nu] \hv_{jk}[\nu]\|^2 \rightarrow \frac{1}{M} \trace ( \Vm^\herm_j[\nu] \Vm_j[\nu] ) = \frac{S_j}{M}. \]
With these limits, we obtain the SINR deterministic approximation (which becomes exact in the large-system regime)
\begin{equation} \label{SINR-mumimo} 
\SINR_{ik}^\mv \rightarrow \frac{  (M -  S_i+1) g_{ik} P_i/S_i}{1 + \sum_{j \in \Ac_{c_i} : j \neq i} m_j g_{jk} P_j}.
\end{equation} 
Notice that, as in the case of single-user beamforming, this limit does not depend on the subcarrier index $\nu$ any longer. 
Also, for the sake of consistency, it is interesting to notice that for $S_i = 1$ 
we recover the expression for single user conjugate beamforming in Equation \eqref{SINR}. 

Under this approximation, the vector of user rates for a given active group of users $\widehat{\Sc}_i \subseteq \Sc_i$ is  given by 
\begin{equation}  \label{mu-mimo-cap}
C_{ik}^\mv(\widehat{\Sc}_i) = \left \{ 
\begin{array}{ll}
0 & \mbox{for} \;\;\; k \in \Sc_i - \widehat{\Sc}_i \\ 
m_i \log \left ( 1 +   \SINR_{ik}^\mv \right ) & \mbox{for} \;\;\; k \in \widehat{\Sc}_i. 
\end{array} \right .
\end{equation}

{\bf Rate region for concentrated MU-MIMO.}
The achievable rate region in the case of concentrated MU-MIMO
is significantly more complicated to express  than for the case of single-user beamforming with 
downlink orthogonal access. In this case, any group of
users $\widehat{\Sc}_i \subseteq \Sc_i$ with size $S_i \leq M$ can be scheduled in the downlink. 
The individual average rate of user $k$ is the convex combination of 
the rates achieved in each group, where the convex combination coefficients depend on the 
downlink scheduling scheme. In particular, we can order 
the groups $\widehat{\Sc}_i \subseteq \Sc_i$ of size $S_i \leq M$ in lexicographic order.  
Let  $\Cm_i^\mv(\widehat{\Sc}_i)$ denote the $1 \times |\Sc_i|$ row vector of user rates 
given in (\ref{mu-mimo-cap}), and let  $\Cm^\mv(\Sc_i)$ denote the $G_i \times |\Sc_i|$  matrix obtained by stacking all the group 
rate rows on top of each other in the same group lexicographic order. 
Define the $1 \times |\Sc_i|$ rate vectors $\Rm^\mv_i$ with components $R^\mv_{ik}$ 
for all users $k \in \Sc_i$.  The rate region obtained by applying TDMA downlink scheduling on top of MU-MIMO for each AP $i$
 is given by the union of all points  $\Rm^\mv_i$ satisfying
\begin{equation} \label{TDMAregion-mu-mimo} 
\Rm^\mv_i \leq \rhov^\transp \Cm^\mv(\Sc_i)
\end{equation}
for some non-negative time-sharing vector $\rhov$ of dimension $G_i \times 1$,  with components satisfying $\sum_{k\in \Sc_i} \rho_k = 1$. 
Notice also that this region generalizes the single-user beamforming TDMA region given in (\ref{TDMAregion}), 
since in this case we have only $|\Sc_i|$ possible groups of size 1, and combining the inequalities
\[ R^\mv_{ik} \leq \rho_{k} C^\mv_{ik} \]
with $\sum_{k \in \Sc_i} \rho_k = 1$ yields again 
Equation (\ref{TDMAregion}). 

Focusing on proportional fairness downlink resource allocation, the AP gives to each user $k \in \Sc_i$ equal air time, such that 
the user rates are given by 
\begin{equation} \label{rate-mu-mimo} 
R_{ik}^\mv = m_i \frac{S_i}{|\Sc_i|}  \log \left ( 1 +   \SINR_{ik}^\mv \right ). 
\end{equation}
The number of downlink streams $S_i$ may be fixed by some technology constraints (e.g., by the number of data streams that the AP chipset is
able to precode), or it may be optimized for each AP, subject to the constraint $S_i \leq \min\{M, |\Sc_i|\}$. In this work, the latter was chosen as we maximize the sum-rate that every AP serves its users individually for every AP.

As before, the average user rate over all possible states of the CSMA/CA CTMC is given by using Equation (\ref{rate-mu-mimo})
in  (\ref{MC-average}).  From the resulting average rates $\{R_{ik}\}$, the throughput CDF can be obtained via Equation (\ref{eqn:cdf}).

%%%%%%%%%%%%%%%%%%%%%%%%%%%%%%%%%%%%%%%%%%%%%%%%%%%
%%%%%%%%%%%%%%%%%%%%%%%%%%%%%%%%%%%%%%%%%%%%%%%%
\subsection{Distributed MU-MIMO}
\label{dmumimo}

In this case we assume that all APs that operate in the same channel cooperate and act as a single virtual mega AP (no contention between the APs for the wireless medium). Therefore, the channel index is not necessary and is dropped for notation simplicity. 

Assume that  $B$ APs with $M$ antennas each form a cluster, pooling together all their $B M$ antennas, 
and collectively serve the population of $K$ UTs with a single antenna each.
This type of distributed MU-MIMO configuration has been widely studied as far as the PHY algorithms and the achievable rates from a communication theoretic viewpoint are concerned (see for example prior work of ours \cite{huh-tulino-caire,massive_mimo_caire,ryan-twc} and others \cite{gesbert,huang_valenzuela}). Recently, experimental results with such systems have been published (see for example our own prior work \cite{airsync_ton} as well as work from others \cite{rahul_jmb,nemox}).   As before, we wish to characterize the 
rate per user in which a subset $\widehat{\Sc}$ of size $S \leq \min \{K, B M\}$ of users is served simultaneously by the distributed MU-MIMO scheme. 
We shall then apply a TDMA scheduling over the user subsets in order to obtain a desired throughput $K$-tuple 
satisfying some required fairness criterion.

As far as the PHY schemes and corresponding performance are concerned, the analysis of 
distributed MU-MIMO is significantly more involved than the previously treated cases 
because the channel vectors from all the AP antennas to any given user are not i.i.d. given the pathloss. 
In fact, letting $g_{ik}$ denote again the channel gain coefficient from AP $i$ to user $k$ and letting $\hv_{ik}[\nu]$ denote the channel small 
fading coefficient,  we have that the composite channel vector is given by 
\[ \hv_k[\nu] = \left [ \begin{array}{c} \sqrt{g_{1k}} \hv_{1k}[\nu] \\ \sqrt{g_{2k}} \hv_{2k}[\nu] \\ \vdots \\ \sqrt{g_{Bk}} \hv_{Bk}[\nu] \end{array} \right ]. \]
For a certain set of $\widehat{\Sc} = \{1,\ldots, S\}$ users to be served, the resulting $B M \times S$ channel matrix takes on the form
\[ \Hm[\nu] = 
\left [ \begin{array}{cccc}
\sqrt{g_{11}} \hv_{11}[\nu] & \sqrt{g_{12}} \hv_{12}[\nu] & \cdots & \sqrt{g_{1S}} \hv_{1S}[\nu]  \\ 
\sqrt{g_{21}} \hv_{21}[\nu] & \sqrt{g_{22}} \hv_{22}[\nu] & \cdots & \sqrt{g_{2S}} \hv_{2S}[\nu]  \\ 
\vdots &  &   &  \vdots \\ 
\sqrt{g_{B1}} \hv_{B1}[\nu] & \sqrt{g_{B2}} \hv_{B2}[\nu] & \cdots & \sqrt{g_{BS}} \hv_{BS}[\nu]   \end{array} \right ], \]
where the effective channel coefficient $\lambda_k$ and the $\SINR$ for user $k \in \widehat{\Sc}$ under joint ZFBF from all the $B$ APs are given by
\begin{equation}\label{sinr_distmumimo}
\lambda_{k}[\nu] = \frac{1}{\left [ \left ( \Hm^\herm[\nu] \Hm[\nu]  \right )^{-1} \right ]_{k,k}}, \quad \SINR_{k}[\nu] = \lambda_k[\nu] \frac{P_{\rm sum}}{S},
\end{equation}
where  we define the sum power of all APs in the cluster as $P_{\rm sum} = \sum_{i=1}^B P_i$ and assume equal power allocation on
all downlink data streams. Note that since all APs operate as a single virtual mega AP, there is no interference between APs and 
thus there is no interference term in the $\SINR$ above.

The problem is that now the structure of the channel matrix depends on the actual set of users $\widehat{\Sc}$ being served. 
Hence, even using asymptotic formulas from random matrix theory, we should evaluate rates for each 
set of ${K \choose S}$ users, for $S = 1,\ldots, \min \{K, BM\}$. With distributed MU-MIMO, relevant numbers are (for example)
$K = 100$, $M = 4$ and $B = 16$. Clearly even enumerating all the possible active user subsets
$\widehat{\Sc}$ is difficult.

Inspired by prior work of ours \cite{huh-tulino-caire} we propose the following simplification: Under certain symmetry conditions in the pathloss coefficients, 
the effective channel coefficient $\lambda_k[\nu]$ for user $k$ under joint ZFBF from all the $B$ APs, in the limit for
large $M$ and large $S$ with fixed ratio $M/S$,  asymptotically takes on the form:
\[ \lambda_k[\nu] \rightarrow \left ( 1 - \frac{S-1}{B M} \right ) \left ( \sum_{i=1}^B g_{ik} \right ). \]
Once more, the limiting expression does not depend on the subcarrier index any longer. 
We use the above expression  as an approximation of the ZFBF effective channel coefficients for general distributed MU-MIMO. 
This has the following appealing interpretation: the distributed MU-MIMO scheme behaves as a concentrated MIMO 
system with pathloss coefficients equal to the sum of the pathloss coefficients and the total number of antennas 
(the number of antennas of all coordinated APs in the cluster).  Thus, we obtain the (approximated)
user downlink rate as:
%\begin{equation}  \label{distributed-mu-mimo-cap}
%C_{k} = \left \{ \begin{array}{l}
%0, \;\;\;\;\; \mbox{for} \;\;\; k \in \Sc - \widehat{\Sc} \\
%\displaystyle{\log \left ( 1 + \left (M - \frac{S - 1}{B} \right ) \left ( \sum_{i=1}^B g_{ik} \right ) \frac{P_{\rm sum}}{S} \right ),} \\ \\
%\;\;\;\;\;\;\;\;\;\;\;\;\;\;\;\;\;\;\;\;\;\;\;\;\; \mbox{for} \;\;\; k \in \widehat{\Sc} \end{array} \right .
%\end{equation}
\begin{equation}  \label{distributed-mu-mimo-cap}
C_{k} = \left \{ \begin{array}{l}
0, \;\; k \in \Sc - \widehat{\Sc} \\
\log \left ( 1 + \left (M - \frac{S - 1}{B} \right ) \left ( \sum_{i=1}^B g_{ik} \right ) \frac{P_{\rm sum}}{S} \right ), \;\; k \in \widehat{\Sc} 
\end{array} \right .
\end{equation}
Notice that this is the effective performance of a virtual concentrated MU-MIMO 
system with ``resource pooling'',  with $M$ effective antennas, effective load (number of active users per antenna) 
equal to $ S/B$,  effective channel gains $g_k = \sum_{i=1}^B g_{ik}$ and effective power $P_{\rm sum}$. 
Notice also that for $B = 1$ the above formula coincides with the case developed for concentrated MU-MIMO 
in the absence of inter-AP interference. 

For a network with $N_a$ APs, letting $B = N_a$ with distributed MU-MIMO yields that the whole network 
corresponds to a single coordination cluster.
In practice, this may be too constraining since other system limitation aspects arise in distributed MU-MIMO architectures. For example, 
since all user data need to be precoded jointly from all the coordinated APs in the cluster, eventually the wired backbone network connecting
the APs will become the system bottleneck. Here we are not concerned with such practical implementation problems of distributed 
MU-MIMO, but for the sake of generality we wish to take into account the case where the network is split into multiple coordination 
clusters. In particular, in some numerical results we consider the case where the network is split into 4 clusters, 
each of which with $B = N_a/4$ APs, and operating on a different 20 MHz channel, such that there is no inter-cluster interference. 
In this case, the same formula (\ref{distributed-mu-mimo-cap}) applies, where we restrict the active user set $\widetilde{\Sc}$ to be a subset of the
users assigned to the given cluster. Similarly, we can cluster the APs in 2 clusters assigning the two 40 MHz non-overlapping channels to each one. User-cluster association can be done through a greedy scheme similar to what we have seen before for the user-AP 
association.

As with the single-user beamforming and concentrated MU-MIMO systems, for distributed MU-MIMO we are interested in 
the per-user downlink throughput under proportional fairness scheduling. 
Letting $S$ denote the number of users simultaneously served by the cluster of APs and assuming that users are given 
equal air time, we have that each user $k$ is served for a fraction of time equal to $\frac{S}{K}$. Hence, 
the proportional fairness scheduler with equal power per stream yields throughput
\begin{equation}\label{rate-dmu-mimo}
R_{k} = \frac{S}{K}  \log \left ( 1 + \left (M - \frac{S - 1}{B} \right ) \left ( \sum_{i=1}^B g_{ik} \right ) \frac{P_{\rm sum}}{S} \right ). 
\end{equation}
These rates can be maximized with respect to the size of the active user set $S = 1,2, \ldots, \min\{K,BM\}$. 
From the resulting average rates $\{R_{ik}\}$, the throughput CDF can be obtained again via Equation (\ref{eqn:cdf}).

\section{Model Validation}
\label{section:validation}

\begin{figure*}
\centering
\begin{subfigure}[t]{.3\textwidth}
   \includegraphics[width=\textwidth]{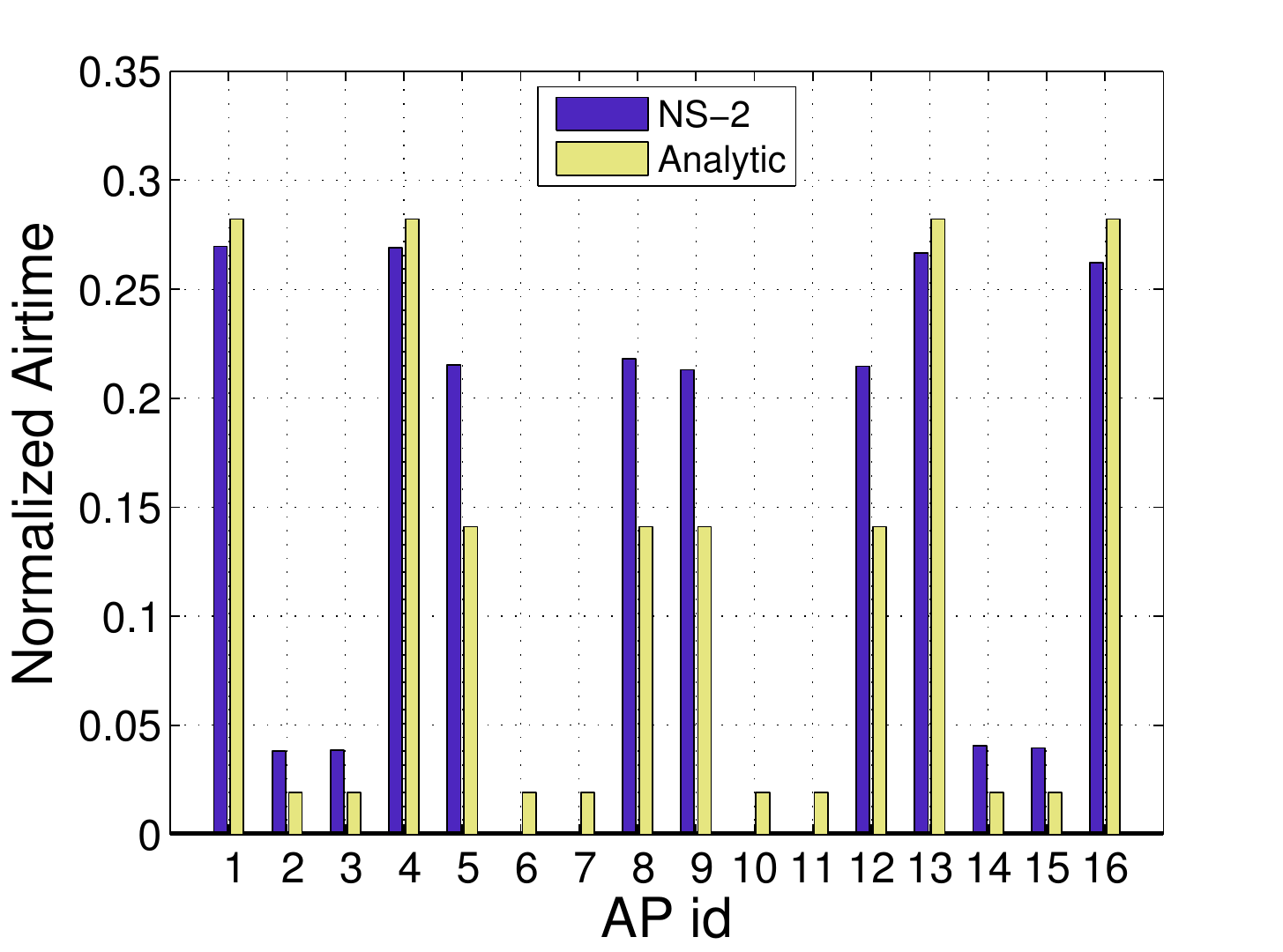} % requires the graphicx package
   \caption{ NS-2 and analytic relative airtime allocations over 16 APs.}
   \label{fig:validation_ns2}
\end{subfigure}
\quad
\begin{subfigure}[t]{.3\textwidth}
   \includegraphics[width=\textwidth]{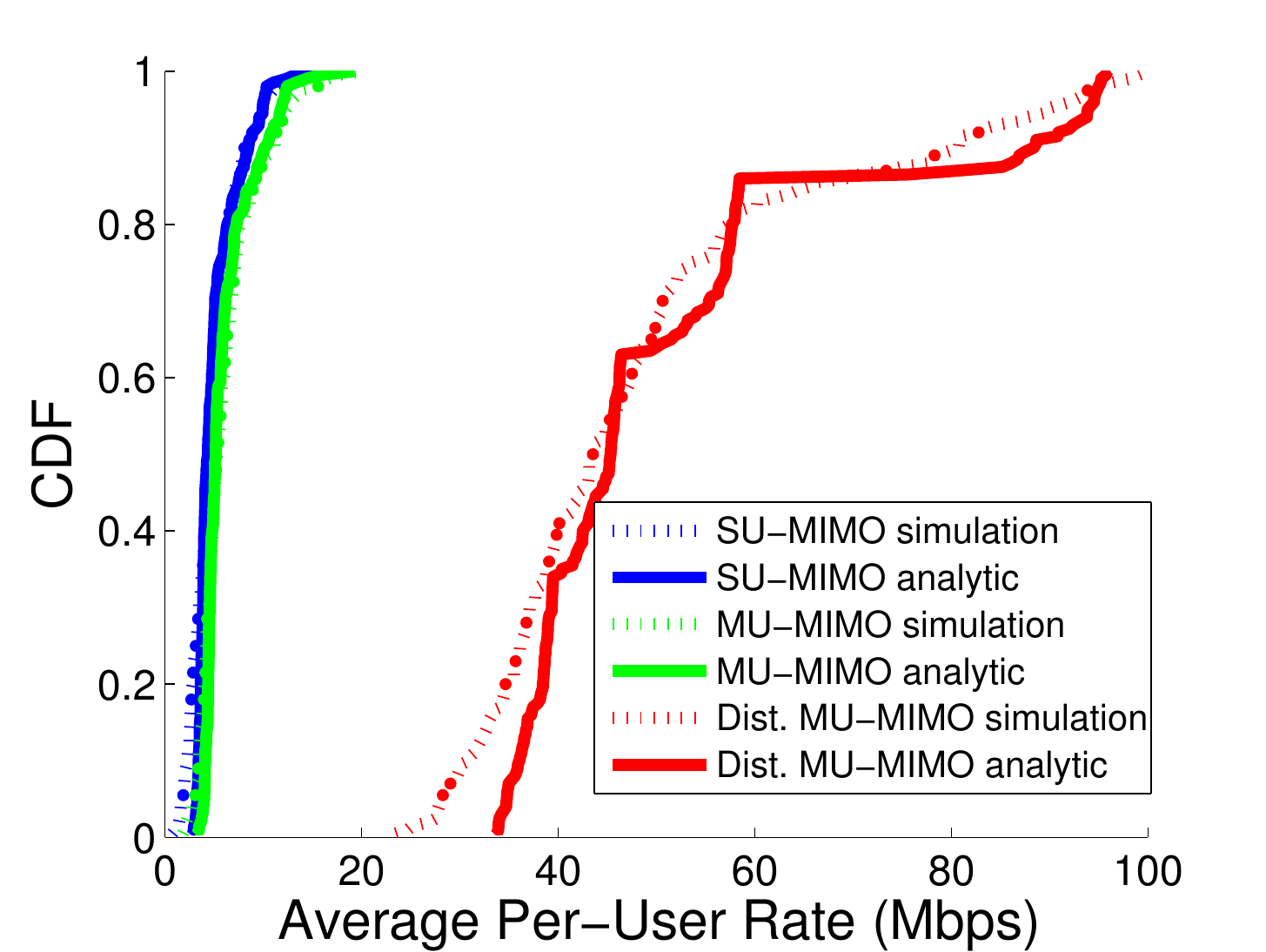} % requires the graphicx package
   \caption{ 20 APs and 4 antennas per AP}
   \label{fig:validation2}
\end{subfigure}
\quad
\begin{subfigure}[t]{.3\textwidth}
   \includegraphics[width=\textwidth]{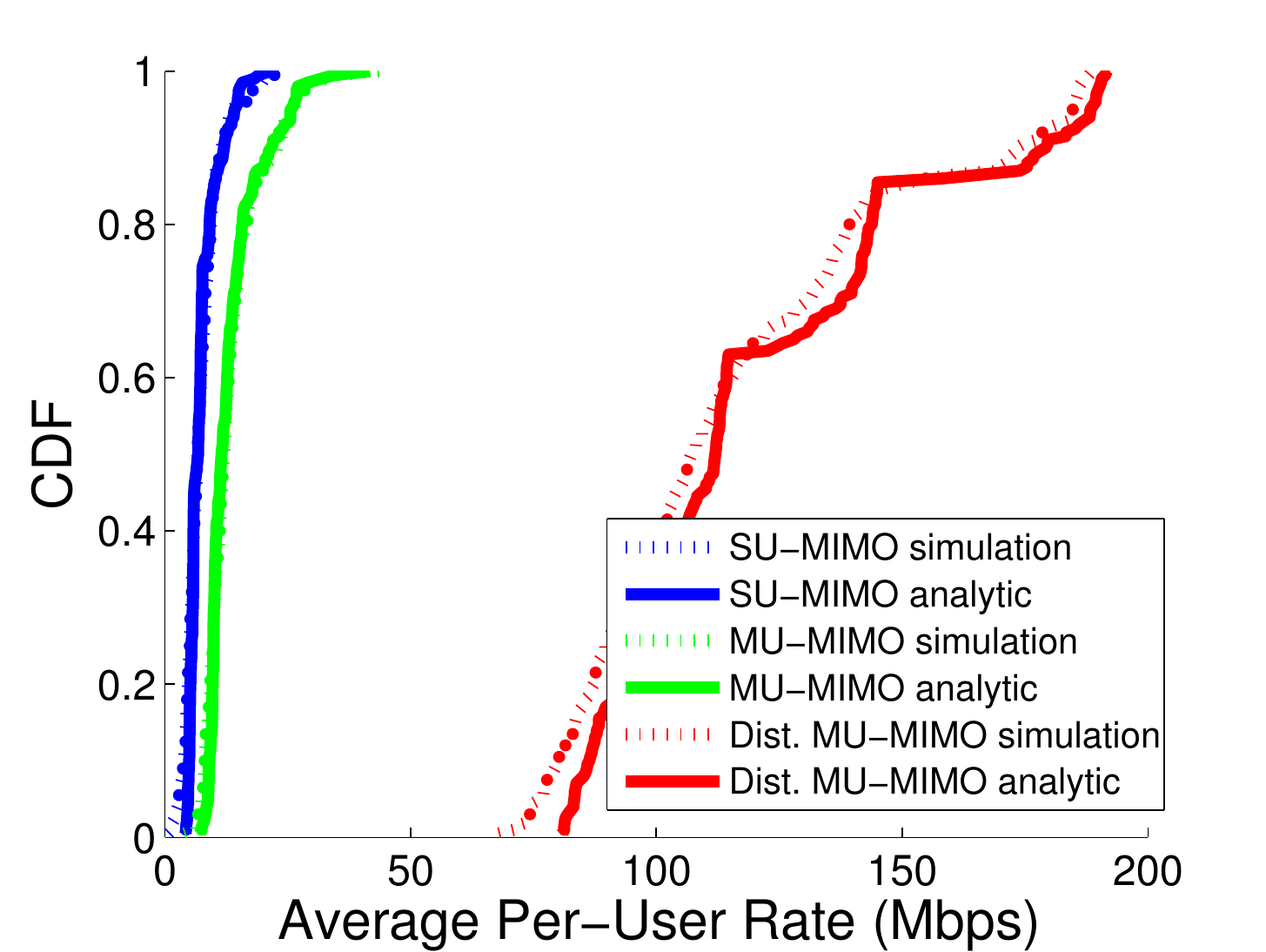} % requires the graphicx package
   \caption{ 20 APs and 10 antennas per AP}
   \label{fig:validation3}
\end{subfigure}
\caption{CDFs and NS-2 results of the avg. user rate for the analytic model vs. simulation for the conference hall scenario.}
\label{fig:conf_hall_validation}
\end{figure*}

\begin{figure*}
\centering
\begin{minipage}[t]{.3\textwidth}
   \includegraphics[width=\textwidth]{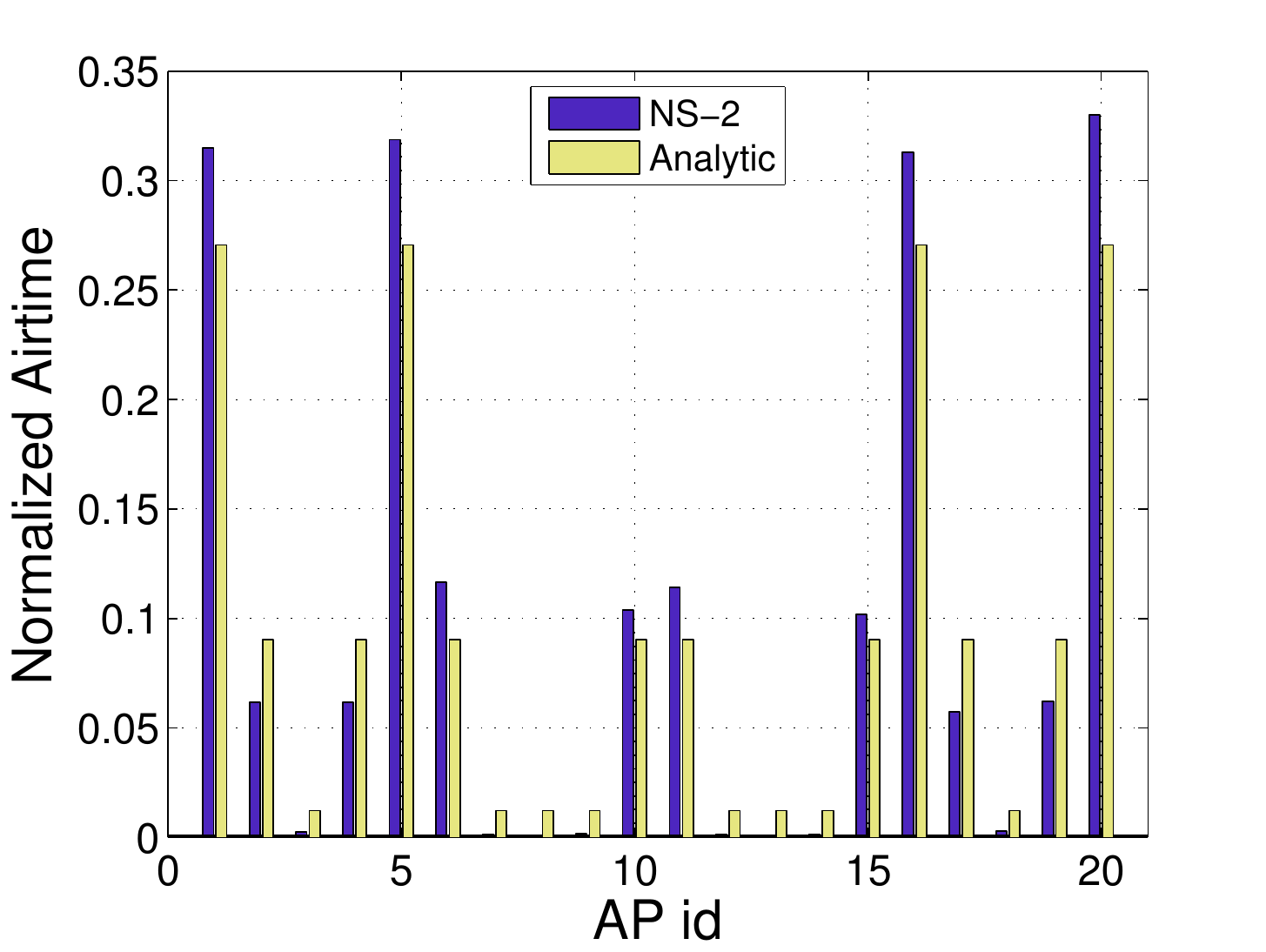} % requires the graphicx package
   \caption{NS-2 and analytic relative airtime allocations over 20 APs.}
   \label{fig:validation_ns22}
\end{minipage}
\quad
\begin{minipage}[t]{.6\textwidth}
\begin{subfigure}[t]{.5\textwidth}
   \includegraphics[width=\textwidth]{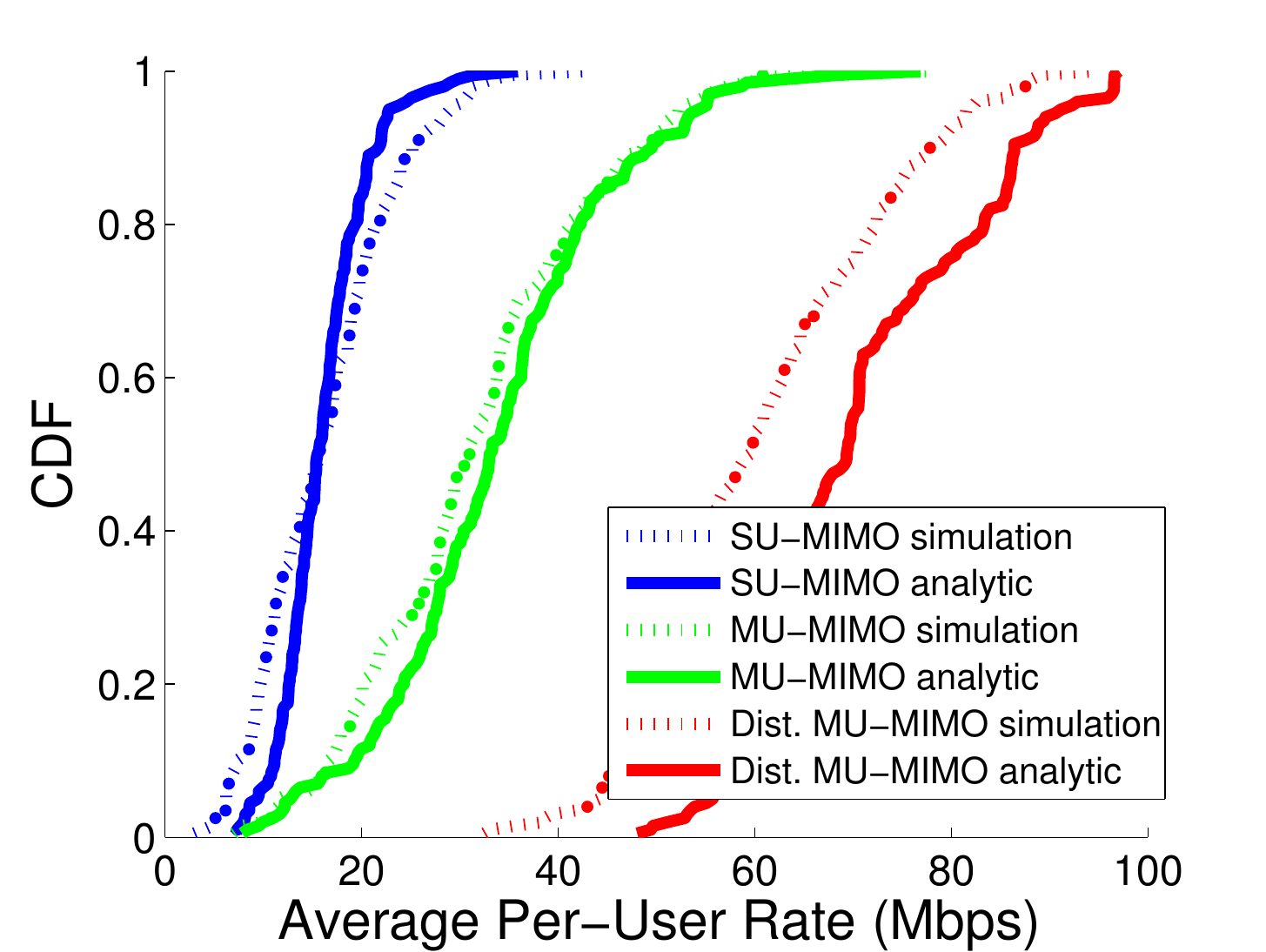} % requires the graphicx package
   \caption{4 rooms, 20 APs, 4 antennas per AP}
   \label{fig:validation4}
\end{subfigure}
\quad
\begin{subfigure}[t]{.5\textwidth}
   \includegraphics[width=\textwidth]{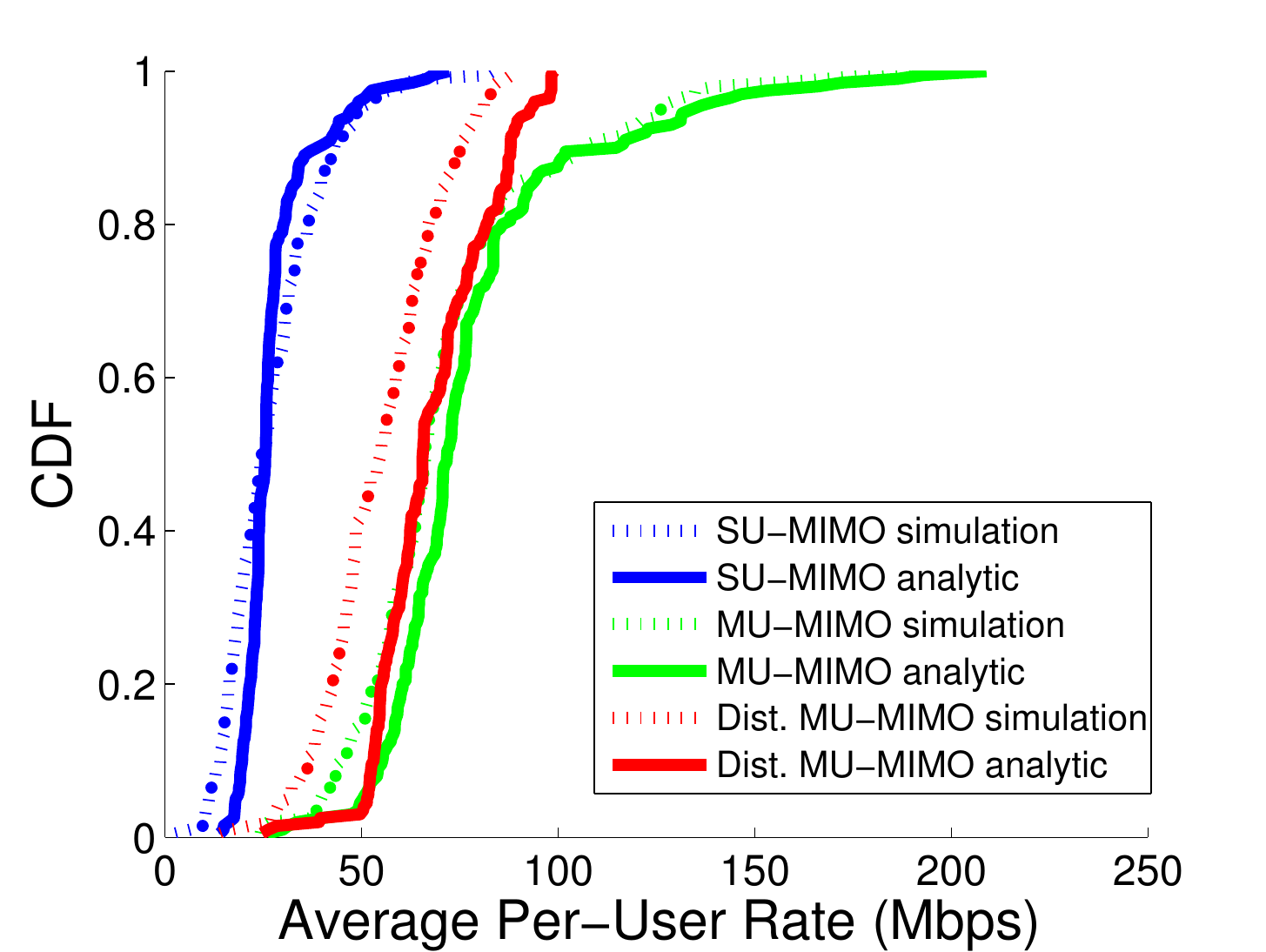} % requires the graphicx package
   \caption{40 rooms, 20 APs, 4 antennas per AP}
   \label{fig:validation6}
\end{subfigure}
\caption{CDFs of the avg. user rate for the analytic model vs. simulation for the office floor scenario.}
\label{fig:office_validation}
\end{minipage}
\end{figure*}
We will proceed with the validation of our model in two steps. First, we will evaluate the accuracy of the analytic PHY models using a custom Monte Carlo 
simulator implemented in Matlab, and then the accuracy of the MAC model (CSMA CTMC) using NS-2 \cite{ns2}. Notice that no open-source 3rd party simulator 
supports the advanced PHY layer schemes that we consider, and thus this two step validation process is required to validate our model in its entirety.

As an example, consider Equation \eqref{MC-average} which gives the rate of a user under single-user
beamforming. This equation uses the rates computed in \eqref{rate-su-mimo} and the stationary distribution of the CSMA CTMC computed in \eqref{ctmc}.
In the first step of the validation process we will use our custom simulator to verify the accuracy of Equation \eqref{rate-su-mimo}, and, in the second step
we will use NS-2 to validate Equation \eqref{ctmc}. A similar process is used for the cases of concentrated MU-MIMO and distributed MU-MIMO, where
in the former case the user rates are given by \eqref{rate-mu-mimo} and in the later by \eqref{rate-dmu-mimo}. 

We start with the first step of the validation process, namely the validation of Equations \eqref{rate-su-mimo}, \eqref{rate-mu-mimo}, 
and \eqref{rate-dmu-mimo}.
In all three cases, we compare the distribution of the average per-user rates as computed by the model 
%(e.g. see Equation \eqref{eqn:cdf}) 
and by our simulation for some representative tractable scenarios of interest, namely an open conference hall (see section \ref{sec:conf_hall}) 
and an office with rooms (see section \ref{sec:office_rooms}).
We utilize the same user locations for both the analytic and simulation results, which are illustrated in Figures \ref{fig:amphi_topo} and \ref{fig:topo_rooms}.
For the scenarios which require APs to use orthogonal channels, the channel selection is performed as in Section \ref{sec:common_params}, where $4$ 
channels of $20$MHz each are shared among the APs. Lastly, the user-AP association is performed as in Section 
\ref{sec:common_params} as well.

Our custom Monte Carlo simulator proceeds as follows. For each topology scenario it creates a large number of channel instantiations. Then, 
for a particular channel realization, it computes the received signal power and the received interference power for each user,
calculates the $\SINR$ of each user under that channel realization, and computes the corresponding (gaussian) user rates.
Finally, it calculates the average achievable rate for each user across all the channel realizations and the empirical distribution
of user rates. It is important to note that the instantaneous $\SINR$ values are computed using a different formula by the Monte Carlo simulator 
depending on the PHY model. For single-user beamforming we use \eqref{SINR}, for concentrated MU-MIMO  \eqref{SINRmumimo}, and
for distributed MU-MIMO \eqref{sinr_distmumimo} (and variations to take into consideration different cluster sizes). Note that these are stochastic 
formulas which depend on the channel realization and it is for this reason that in Section \ref{section:model} we obtain simple yet accurate 
deterministic formulas which form the gist of our analytical model, e.g. Equation \eqref{SINR-mumimo} in place of \eqref{SINRmumimo}.
Notice also that our deterministic formulas remove other stochastic aspects of the problem as well, like the dependence of user MIMO 
scheduling in instantaneous $\SINR$ values.

For the SU-MISO simulations, each transmitter transmits to a random user located within its cell on each iteration.  The transmitters are assumed 
to have channel state information, so the transmitters beamform to their selected user.  Physically adjacent APs operate on separate channels, 
such that users experience interference only from non-neighbor APs transmitting on the same frequency.  Each AP transmits with power 
$P=90\text{dB}$ above the noise floor, and at a bandwidth of 20 MHz.  Including pathloss, the users receive their respective signals at 
typical Wi-Fi SNRs in the range of roughly 0-40dB.

In the case of the .11ac-like approach, APs retain the same channel allocation as in the SU-MISO simulations, such that neighboring 
APs operate on different frequencies.  Instead of transmitting to a single user, however, they beamform to a random subset of the users in 
each cell using ZFBF.  Based on simulation trials, we determined the optimal number of users (on average) to serve simultaneously in 
each cell, which was typically 2 or 3.  Again, each AP transmits with power $P=90\text{dB}$ above the noise floor at a bandwidth of 
20 MHz. 

The distributed MIMO simulations utilize a clustering approach in which groups of 5 APs achieve sufficient synchronization such that they 
act as a single 20-antenna AP.  In this case, all interference is nulled within a given cluster, and each cluster is assigned to a channel 
as in the previous two simulations.  We assume that the APs pool their power, such that total transmit 
power is $\sim 97\text{dB}$ within each cluster.  This simplifies the treatment of the precoder calculation, but it may allow individual 
APs to transmit higher than $90\text{dB}$ power 
above the noise floor.  In practice, we would need to utilize results on ZFBF under per-antenna power constraints \cite{caire-huh-per-antenna}, 
though we avoid this issue since it is not treated in the model.  As in the previous simulations, each cluster occupies 20 MHz of bandwidth.

Figures \ref{fig:validation2}, \ref{fig:validation3} and \ref{fig:office_validation} display the CDFs of the average per-user rates for the simulations and 
the models.  
%It is evident that as the number of antennas per AP increases, the accuracy of our model tends to match the simulation curves.
%Notice that if the number of users increases with the number of AP antennas, the 
%accuracy of the asymptotic model is even greater (Figures \ref{fig:validation1} and \ref{fig:validation4}).
Although the analytic models make a number of assumptions, for example, in the single-user beamforming case that the number of antennas $M$
per AP is large, in the concentrated MU-MIMO case that the number of antennas and users per AP is large, and in the distributed MU-MIMO case 
that, in addition, there is a specific geometric symmetry in the topology, we see that the Monte Carlo simulation results and the analytical results 
are quite close.

Lastly, we proceed to the second step of the validation process and evaluate the accuracy of the CSMA CTMC model used in this work for the open conference 
hall scenario depicted in Figure \ref{fig:amphi_topo} using NS-2 \cite{ns2}. In order to infer the APs' relative airtime in NS-2, an 802.11 scenario was 
constructed where every AP was transmitting a UDP stream with constant bit rate (CBR) traffic of $7$ Mbps to a closely located UT. The basic rate was set to $1$ Mbps, the data rate to $11$ Mbps, the capture future was turned of and the carrier sensing and receiving thresholds were arranged so that the same interference landscape was created for both the simulation and the CTMC model. Every AP in isolation could transmit at a rate of $4.97$ Mbps, and thus, by simulating the achievable throughput when all APs were transmitting, we can estimate the relative airtime of each AP. The 
results can been seen in Figures \ref{fig:validation_ns2} and \ref{fig:validation_ns22} for 16 and 20 APs respectively. Note that a similar CTMC CSMA model has been validated under small scale scenarios in 
\cite{csma_boe}.
%\begin{figure}
%\centering
%   \includegraphics[width=0.4\textwidth]{ns2_analytic} % requires the graphicx package
%   \caption{NS-2 and analytic relative airtime allocations over the APs for the open conference hall}
%   \label{fig:validation_ns2}
%\end{figure}

\section{Results}
\label{section:simulations}
In the following section, results and insights from the application of the analytic model in various practical scenarios will be presented. 
\subsection{OFDM and MCS overhead} 
\label{sec:ofdmoverhead}
We discuss the rate degradation from having actual quantized rates based on the 802.11 standard's Modulation and Coding schemes (MCS) along with the overhead incurred by OFDM. 
There are of course other sources of overhead as well, e.g. the CSMA overhead which been included in our model already (see Section \ref{sec:macprimer}), and the overhead to collect 
channel state information and/or coordinate remote APs in the context of distributed MU-MIMO which will be discussed later (see Section \ref{sec:appendix}). 

Starting from the OFDM overhead we note that based on the 802.11ac standard \cite{802.11ac} for channels of $20$ MHz bandwidth, only $52$ of the $64$ subcarriers carry data (the 
rest are devoted to pilots or are nulled). 
For 40MHz and 80MHz channels the corresponding numbers are $108$ out of $128$ subcarriers, and $234$ out of $256$ subcarriers respectively (notice that these overheads slightly 
vary for 802.11n and 802.11ac, the 
numbers of the latter are adopted for reasons of simplicity). Finally, the OFDM Cyclic Prefix (CP) is included in the Guard Interval (GI) that is prepended to every OFDM symbol for a total 
overhead of an extra $20\%$ 
(assuming a normal GI of $0.8$ $\mu$s with total symbol duration of $4$ $\mu$s).

As a more realistic approach to the rate calculation, we map the received SINRs into a discrete set of modulation and coding pairs. While we acknowledge that choosing the best among 
several discrete modulation and coding 
options (known as rate adaptation) is non-trivial, for the sake of simplicity we assume that we can choose the best scheme based on the received SINR. Table \ref{tbl:acm} provides
one such mapping that corresponds to the 9 mandatory MCSs of 802.11ac
\cite{802.11ac}, keeping in mind that mappings vary by vendor or may
be dynamically chosen in practical scenarios.

The scenarios presented in the rest of the section cover typical yet interesting cases of WiFi deployments, along with more challenging situations. Specifically: i) a densely populated 
conference hall, ii) an open-floor plan office space, iii) an office building floor with separated rooms and iv) an event stadium with high capacity. For all scenarios 4 antennas per AP are 
assumed (the maximum number allowed by the 802.11n standard and supported by 802.11ac implementations to date) and the user receivers are equipped with 1 antenna, as is the 
typical case for handheld devices.

\begin{table}
  \begin{center}
    \begin{tabular}{| c | c | c | c |}
    \hline
    802.11ac MCS Index & Modulation & Code Rate & SNR  \\ \hline
    0 & BPSK & 1/2 & $\geq$ 2dB \\ \hline
    1 & QPSK & 1/2 & $\geq$ 5dB \\ \hline
    2 & QPSK & 3/4 & $\geq$ 8dB \\ \hline
    3 & 16-QAM & 1/2 & $\geq$ 12dB \\ \hline
    4 & 16-QAM & 3/4 & $\geq$ 15dB \\ \hline
    5 & 64-QAM & 2/3 & $\geq$ 18dB \\ \hline
    6 & 64-QAM & 3/4 & $\geq$ 21dB \\ \hline
    7 & 64-QAM & 5/6 & $\geq$ 24dB \\ \hline
	8 & 256-QAM & 3/4 & $\geq$ 27dB \\
    \hline
    \end{tabular}
  \end{center}
  \caption{Modulation/Coding pairs from IEEE 802.11ac and the corresponding SINRs at which they can be selected.}
  \label{tbl:acm}
\end{table}

\subsection{Conference Hall}
\label{sec:conf_hall}
%The plots show the average user throughput for the scenarios of a CSMA system and a system that doesn't rely on CSMA and allows all APs to transmit concurrently by increasing the
%CCA (Clear Channel Assessment) threshold for CSMA.

\begin{figure*}
\centering
\begin{subfigure}{.3\textwidth}
\centering
   \includegraphics[width=.8\textwidth]{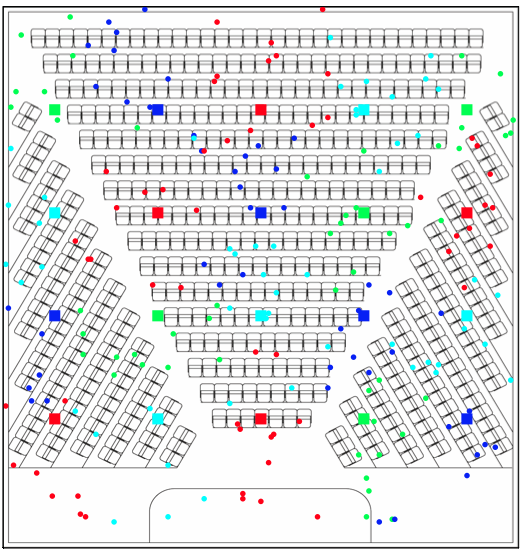} % requires the graphicx package
   \caption{Conference hall with 20 APs and 200 users.}
   \label{fig:amphi_topo}
\end{subfigure}
\begin{minipage}{0.3\textwidth}
\begin{subfigure}{\textwidth}
   \includegraphics[width=\textwidth]{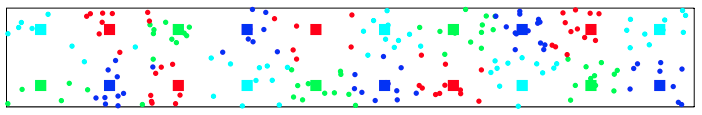} % requires the graphicx package
   \caption{Open-floor office building with 20 APs and 200 users.}
   \label{fig:topo_openfloor}
\end{subfigure}
\quad
\quad
\begin{subfigure}{\textwidth}
   \includegraphics[width=\textwidth]{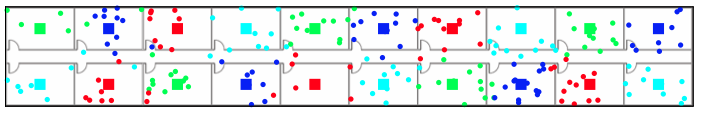} % requires the graphicx package
   \caption{Office building with rooms with 20 APs and 200 users.}
   \label{fig:topo_rooms}
\end{subfigure}
\end{minipage}
\begin{subfigure}{.3\textwidth}
   \includegraphics[width=\textwidth]{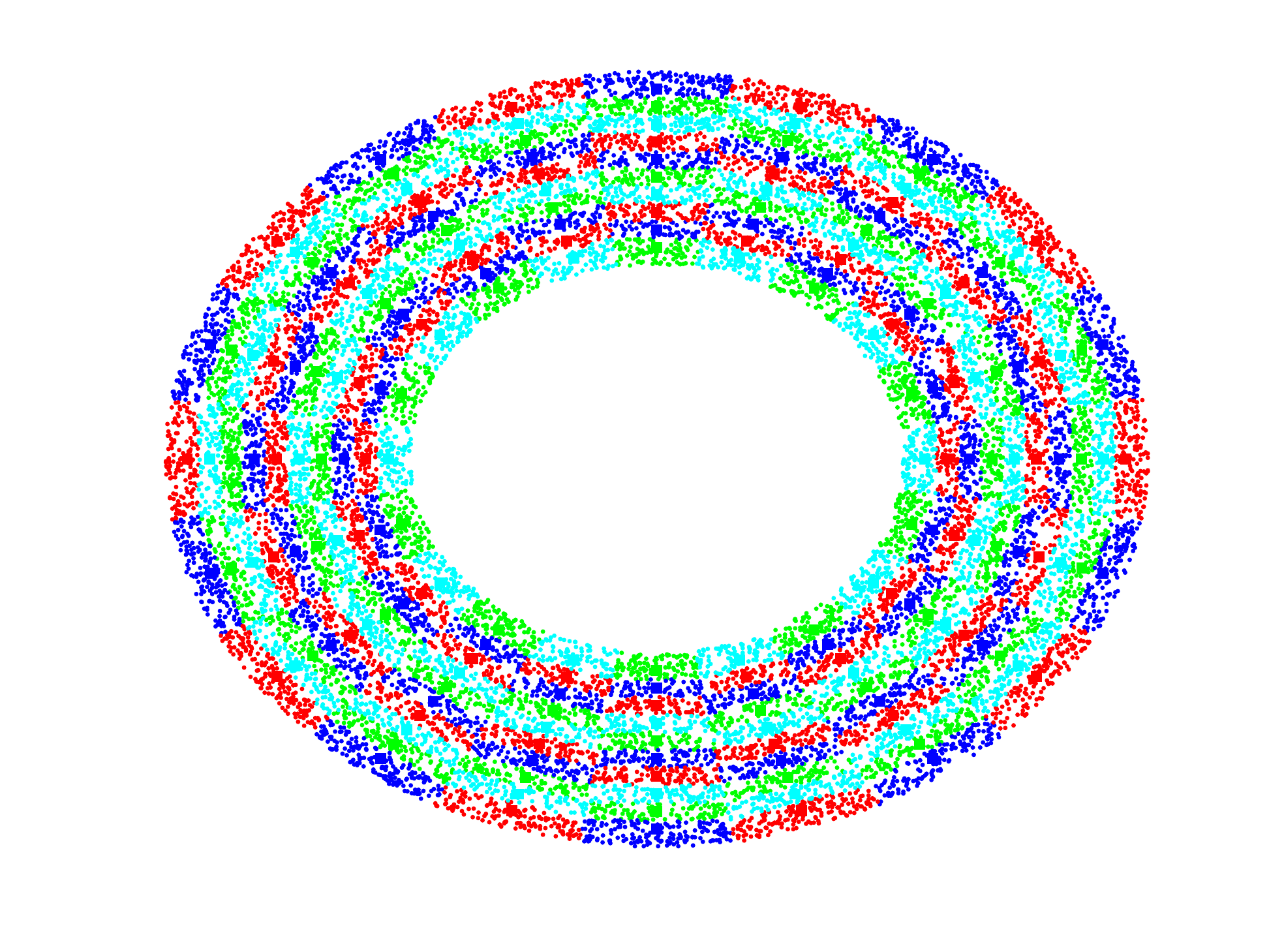} % requires the graphicx package
   \captionof{figure}{ Stadium scenario with 200 APs and 20000 users.}
   \label{fig:stadium_topo}
\end{subfigure}
\caption{ AP placement, channel allocation and user-to-AP assignment for the scenarios examined.}
\end{figure*}

\begin{figure*}
   \centering
   \begin{subfigure}[t]{0.3\textwidth}
   \centering
   \includegraphics[width=\textwidth]{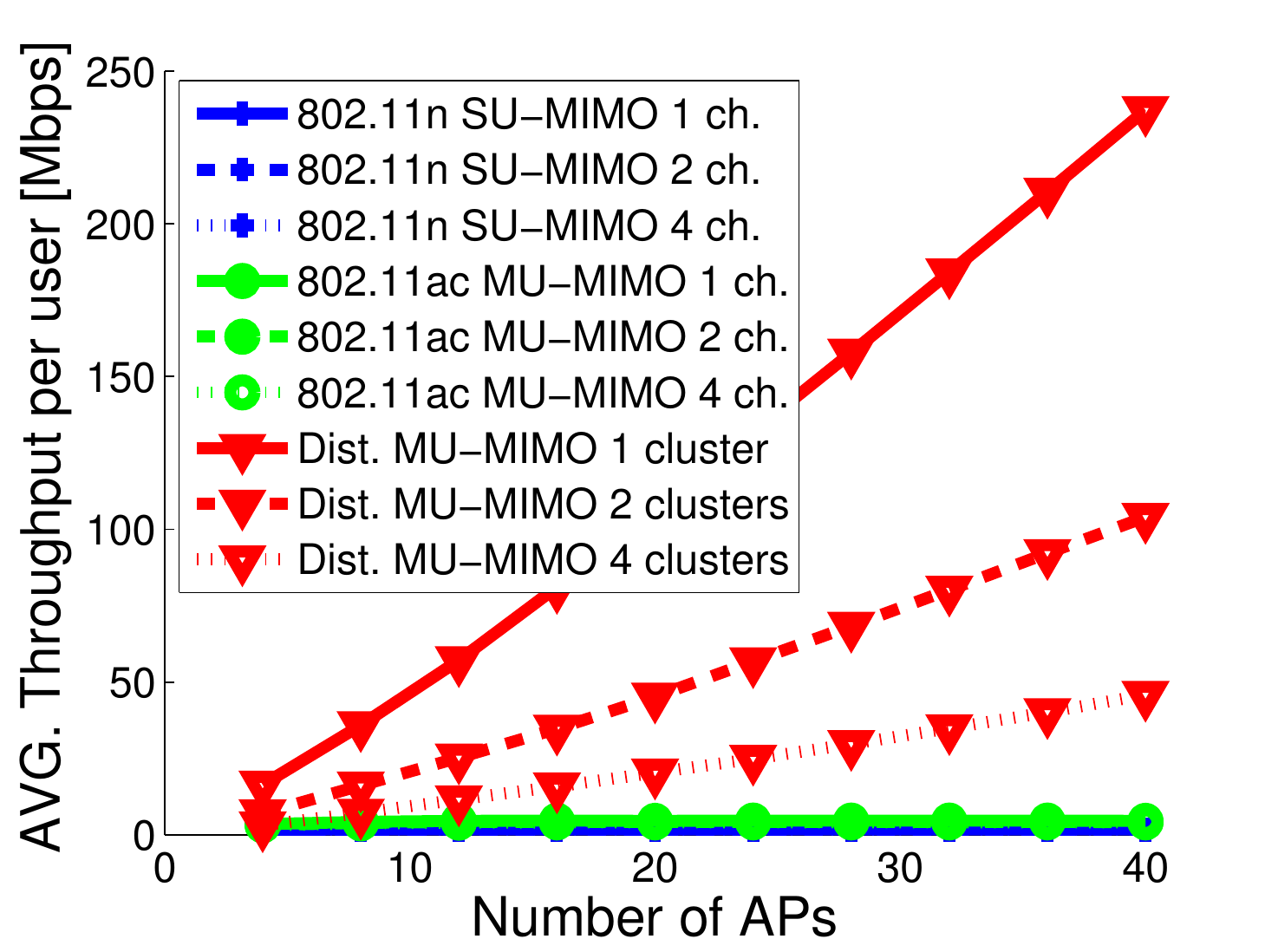} % requires the graphicx package
   \caption{All technologies with $\text{CCA}=10$dB}
   \label{fig:amphi1}
   \end{subfigure}
   \quad
   \begin{subfigure}[t]{0.3\textwidth}
   \includegraphics[width=\textwidth]{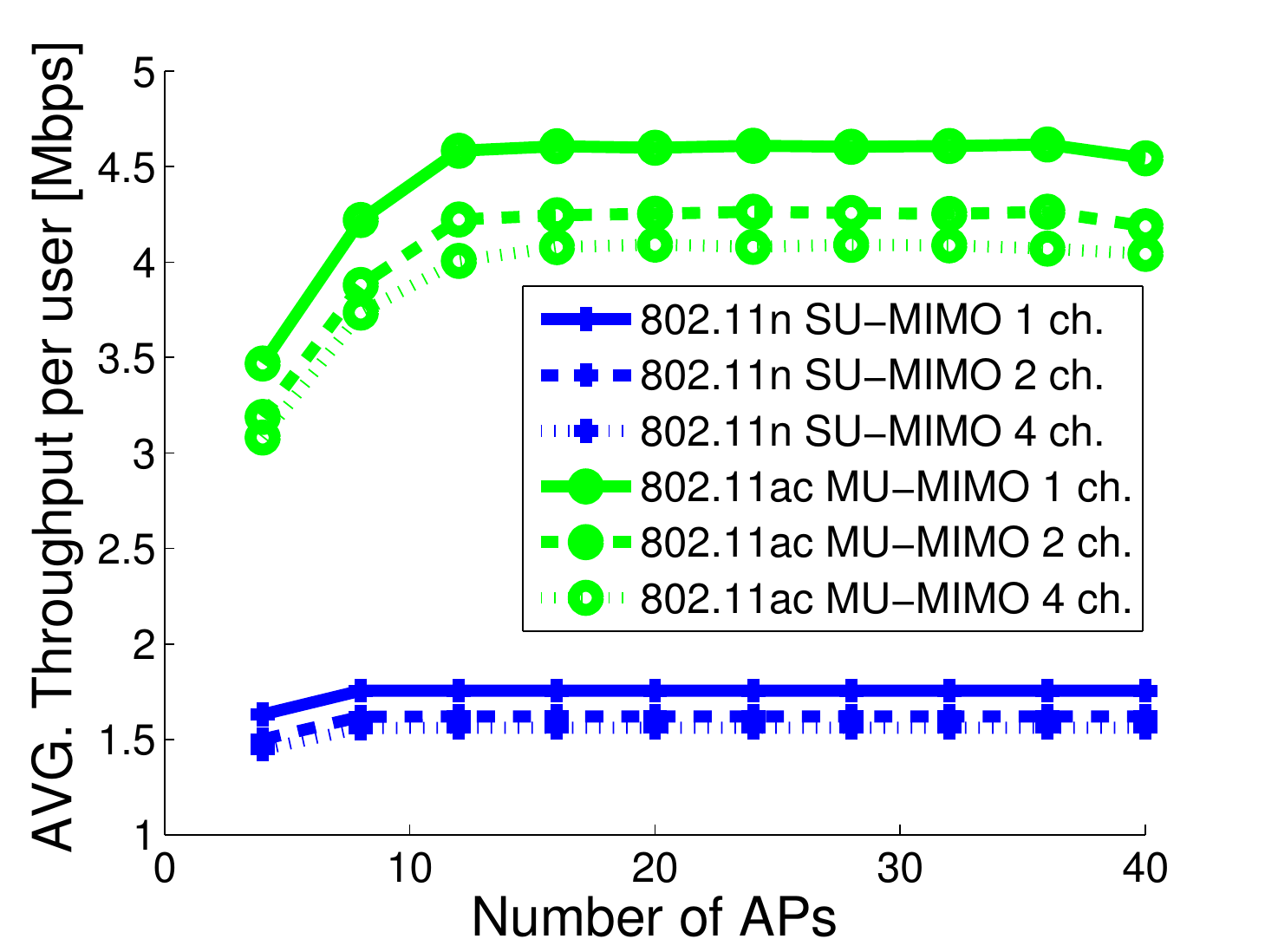} % requires the graphicx package
   \caption{ 802.11n and 802.11ac with $\text{CCA}=10$dB}
   \label{fig:amphi2}
   \end{subfigure}
   \quad
   \quad
   \begin{subfigure}[t]{0.3\textwidth}
   \includegraphics[width=\textwidth]{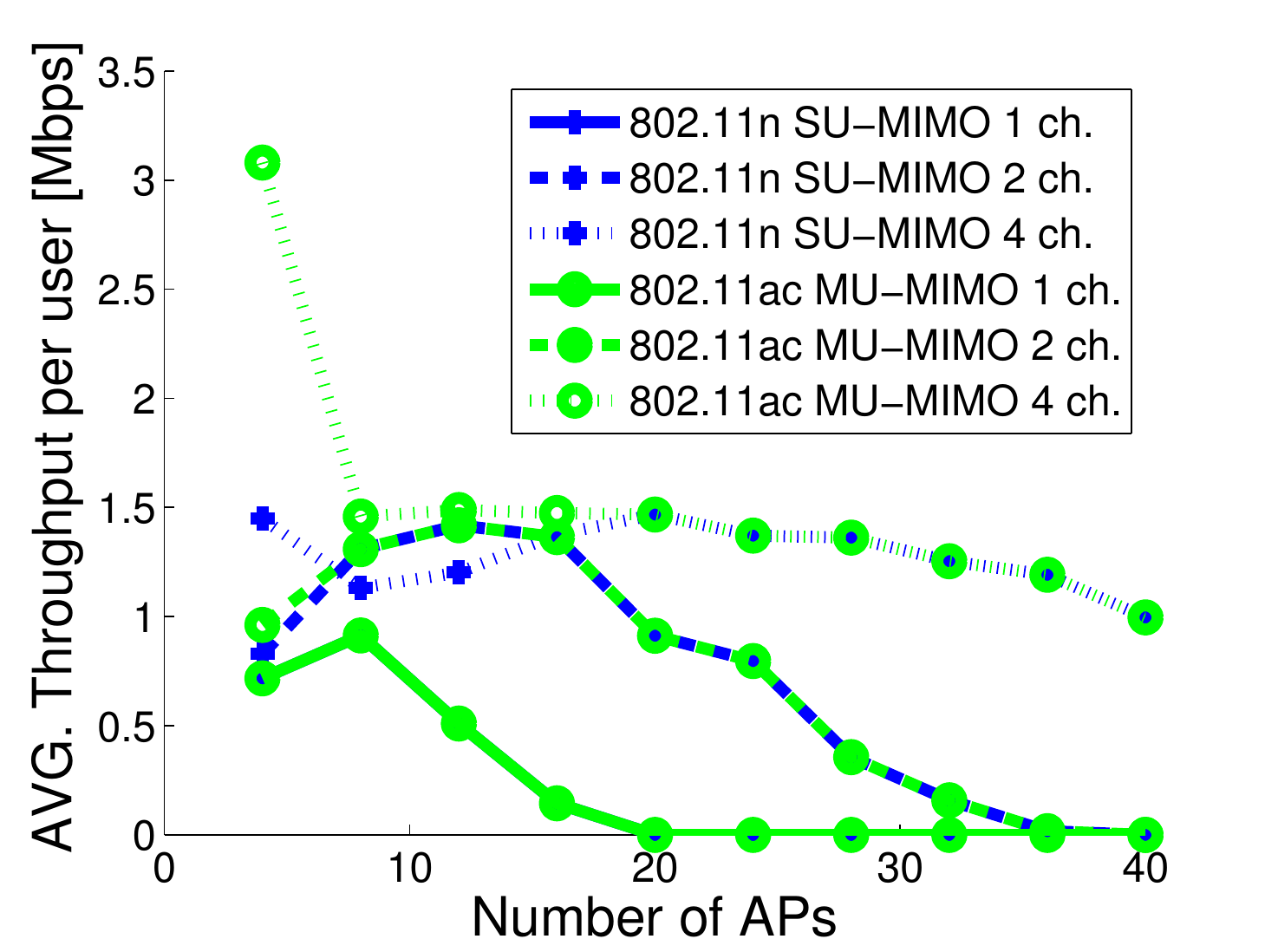} % requires the graphicx package
   \caption{ 802.11n and 802.11ac with no CCA}
   \label{fig:amphi3}
   \end{subfigure}
    \caption{Average throughput per user in a conference hall with $200$ users varying CCA thresholds and 90dB transmit power.}
\end{figure*}

A conference hall of dimension $20$mx$20$m with $200$ users and a varying number of APs (see Figure \ref{fig:amphi_topo}) is examined. 
%{\bf Increasing the number of APs}:\\
 Figure \ref{fig:amphi1} plots the average throughput per user against the number of APs. It is obvious that in this scenario, a coordinated solution where interference is suppressed 
through the distributed MU-MIMO system is 
highly favorable. As the number of APs increases, the average user throughput increases for the distributed MU-MIMO technology. The greatest gains come if we utilize the whole 
bandwidth as a single channel and let both 
APs and user receivers transmit and receive in the whole $80$MHz band. Nevertheless, even in the case where we cluster neighboring APs to act as a single AP, assign a $20$MHz 
channel to each cluster, and assign each 
user to a single cluster (see Section \ref{dmumimo}) the gains are still enormous compared to the non-coordinated technologies. 
%We should mention here that the distributed MU-MIMO clustering approach is not the only way to implement such a technology, but it is a rather natural one.In fact, provided that the 
%APs are able to transmit in the whole 
%$80$MHz another approach would be to have an AP transmit in more than one channels, if users can only receive at smaller channel bandwidths. Such an approach yields similar
%results with the single cluster approach since 
%users would be multiplexed in time and frequency instead of only time in the latter case. 

It is notable that even under the idealized CSMA scenario where collisions are assumed to be insignificant, the throughputs of 802.11n SU-MISO and 802.11ac MU-MIMO are quickly 
saturated. Thus, increasing the number of APs does not provide any extra gains as can be seen in Figure \ref{fig:amphi2}. 
Also, the small gains in the figure that different channelization options provide to non-coordinated systems under a strict CCA threshold are due to the reduced OFDM overheads when a 
smaller number of larger channels is used (see Section \ref{sec:ofdmoverhead}). Notice that this is only true under the assumption of no collisions for the idealized CSMA model. If 
collisions are taken into account frequency re-use has a greater impact on the throughput. 

\begin{figure*}[!ht]
\centering
\begin{subfigure}[t]{0.3\textwidth}
   \includegraphics[width=\textwidth]{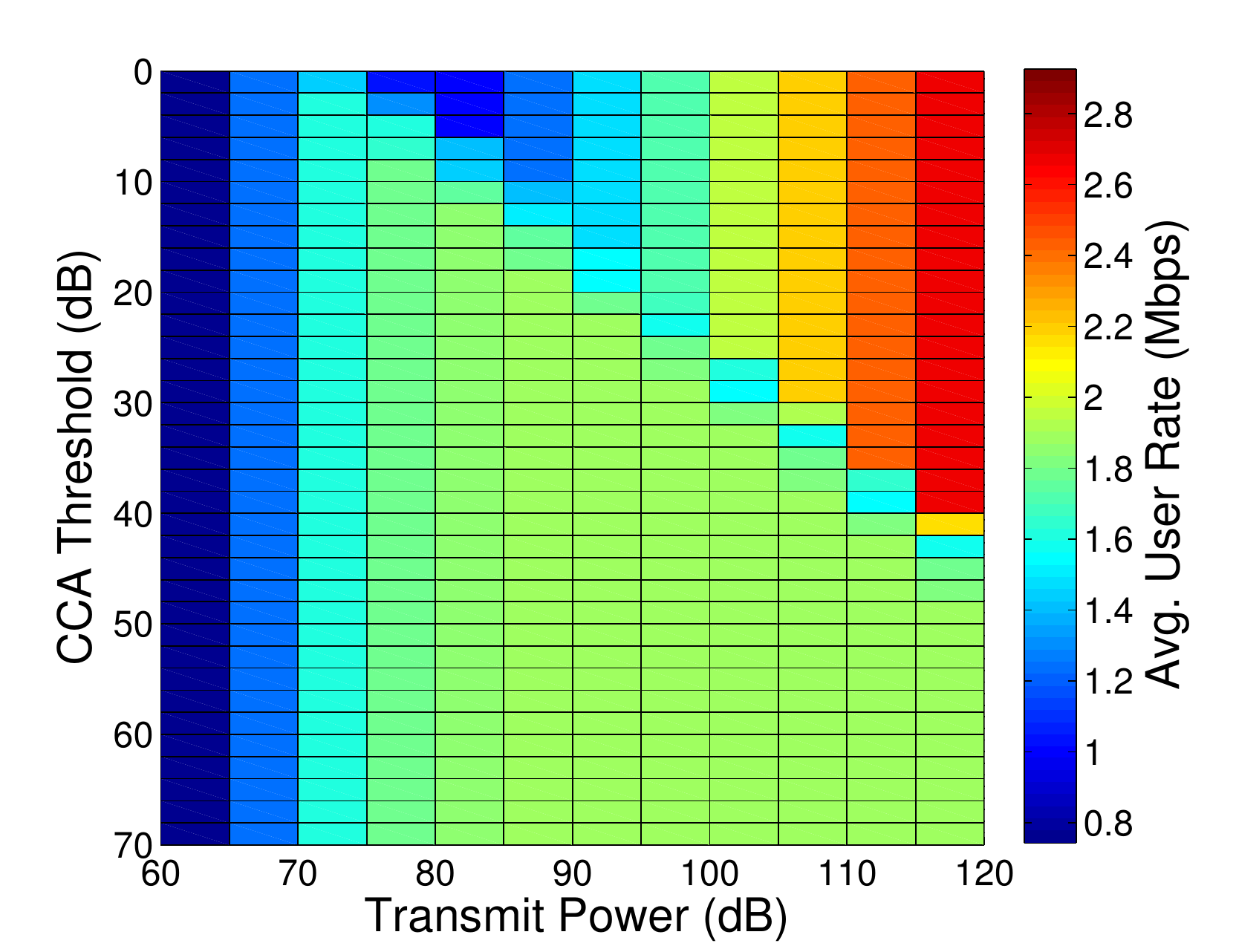} % requires the graphicx package
   \caption{802.11n SU-MISO}
   \label{fig:cca1}
\end{subfigure}
\quad
\begin{subfigure}[t]{0.3\textwidth}
   \includegraphics[width=\textwidth]{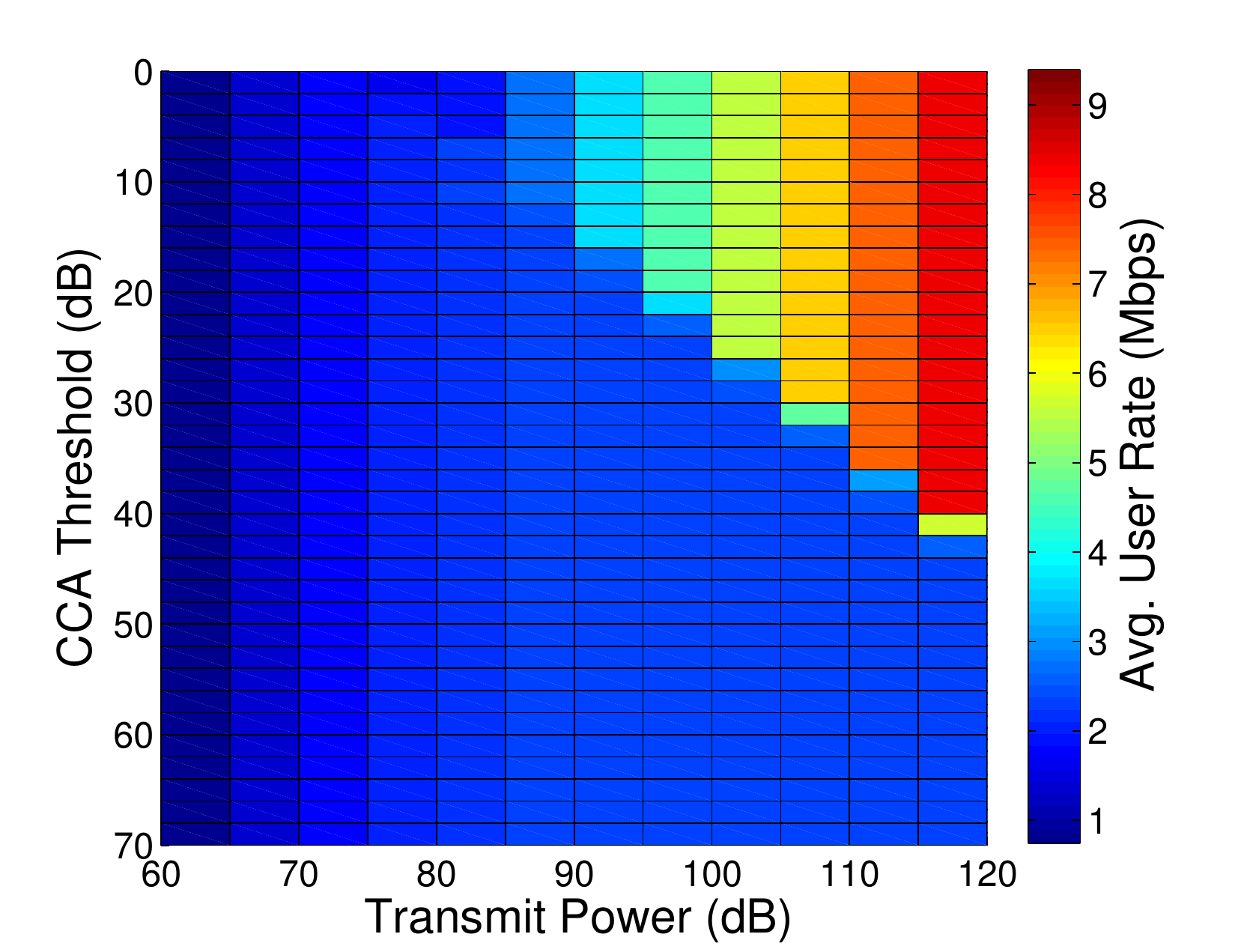} % requires the graphicx package
   \caption{802.11ac MU-MIMO}
   \label{fig:cca2}
\end{subfigure}
\quad
\begin{subfigure}[t]{0.3\textwidth}
   \includegraphics[width=\textwidth]{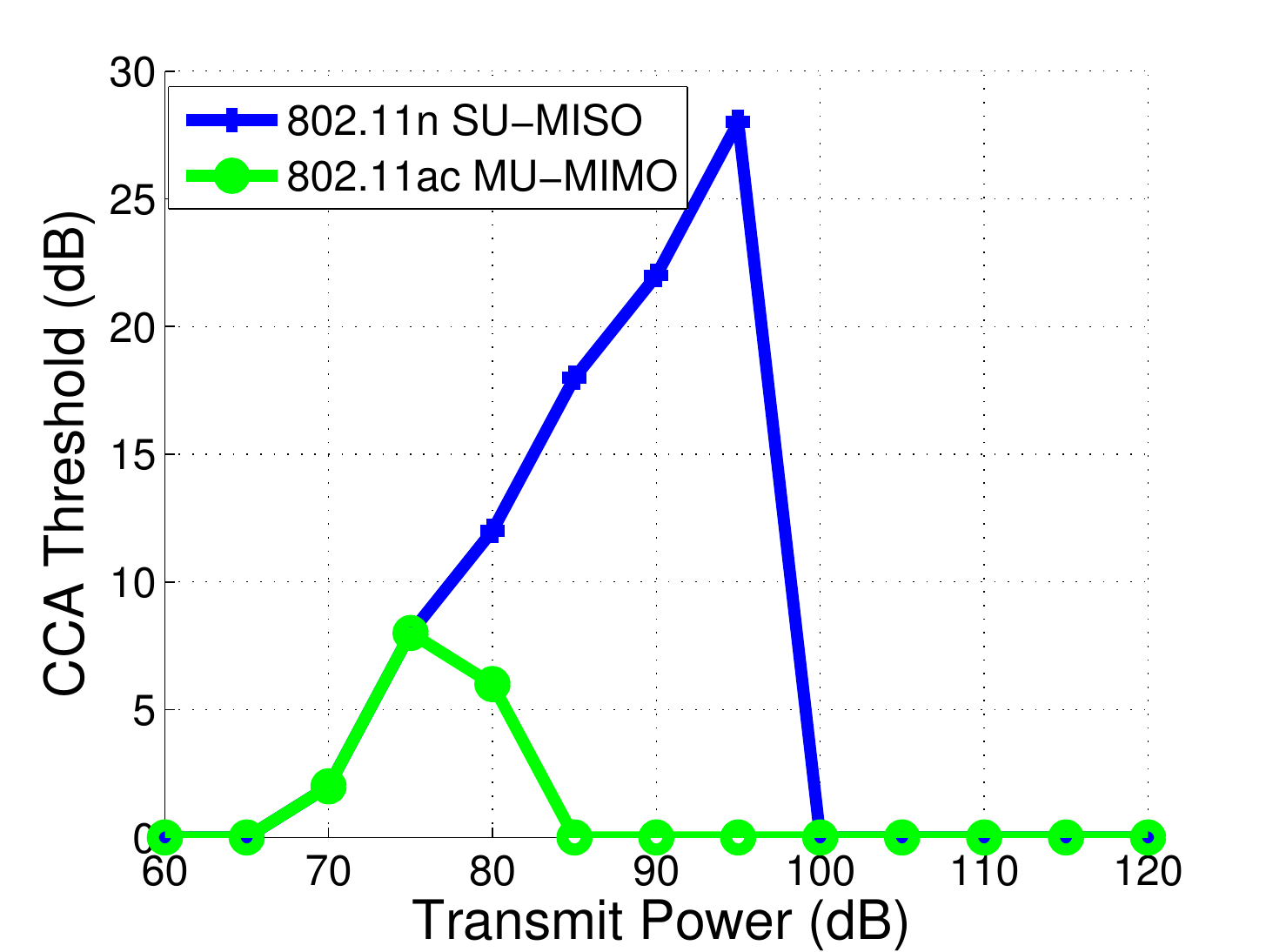} % requires the graphicx package
   \caption{ Minimum optimal CCA threshold for a range of transmit powers from 60 to 120 dB}
   \label{fig:cca3}
\end{subfigure}
\caption{CCA threshold and transmit power optimization for 802.11n SU-MISO and 802.11ac MU-MIMO.}
\end{figure*}

{\bf CSMA impact as number of APs increases:}
In Figures \ref{fig:amphi2} and \ref{fig:amphi3} we see the average per user throughput for different channelization options for the case of CSMA with the default CCA threshold of 10dB 
above the noise floor and an unbounded 
CCA threshold (no CSMA present). It is worth to notice that 802.11ac MU-MIMO, being more susceptible to interference, performs significantly worse under the regime of no CSMA. 
It performs similar to 802.11n's robust 
conjugate beamforming gains as the number of interferers grows (Figure \ref{fig:amphi3}). Indeed, in this interference limited regime, the MU-MIMO system selects only one user to 
be served at a time, reverting to a SU-
MISO scheme. On the other hand, as can be seen in Figure \ref{fig:amphi2}, using CSMA random access clearly benefits the MU-MIMO scheme as the multiplexing gains are larger.

{\bf CCA Threshold and Power Control:} Figures \ref{fig:cca1} and \ref{fig:cca2} show the change in the average user throughput as power increases and the CCA (Clear Channel 
Assessment) threshold is changed for 802.11n SU-MISO and 802.11ac MU-MIMO. For this optimization, Gaussian rates have been assumed instead of the discretized rate allocation we 
introduced. Also, a $4$ channel system with $20$ MHz per channel and 20 APs serving 200 users is assumed for these computations. We can see from these plots that SU-MISO is more robust to interference as the 
power of all APs increases, whereas for MU-MIMO to perform with a noticeable multiplexing gain a lower CCA threshold has to be chosen. 
In Figure \ref{fig:cca3} we see the optimal CCA threshold for a range of transmit powers from 60 to 120 dB for the two non-coordinated technologies.
Notice that thanks to the model we have introduced, such an optimization can be quickly performed and trade-offs can be efficiently explored. 

\begin{figure*}
   \centering
   \begin{minipage}[t]{.3\textwidth}
\includegraphics[width=\textwidth]{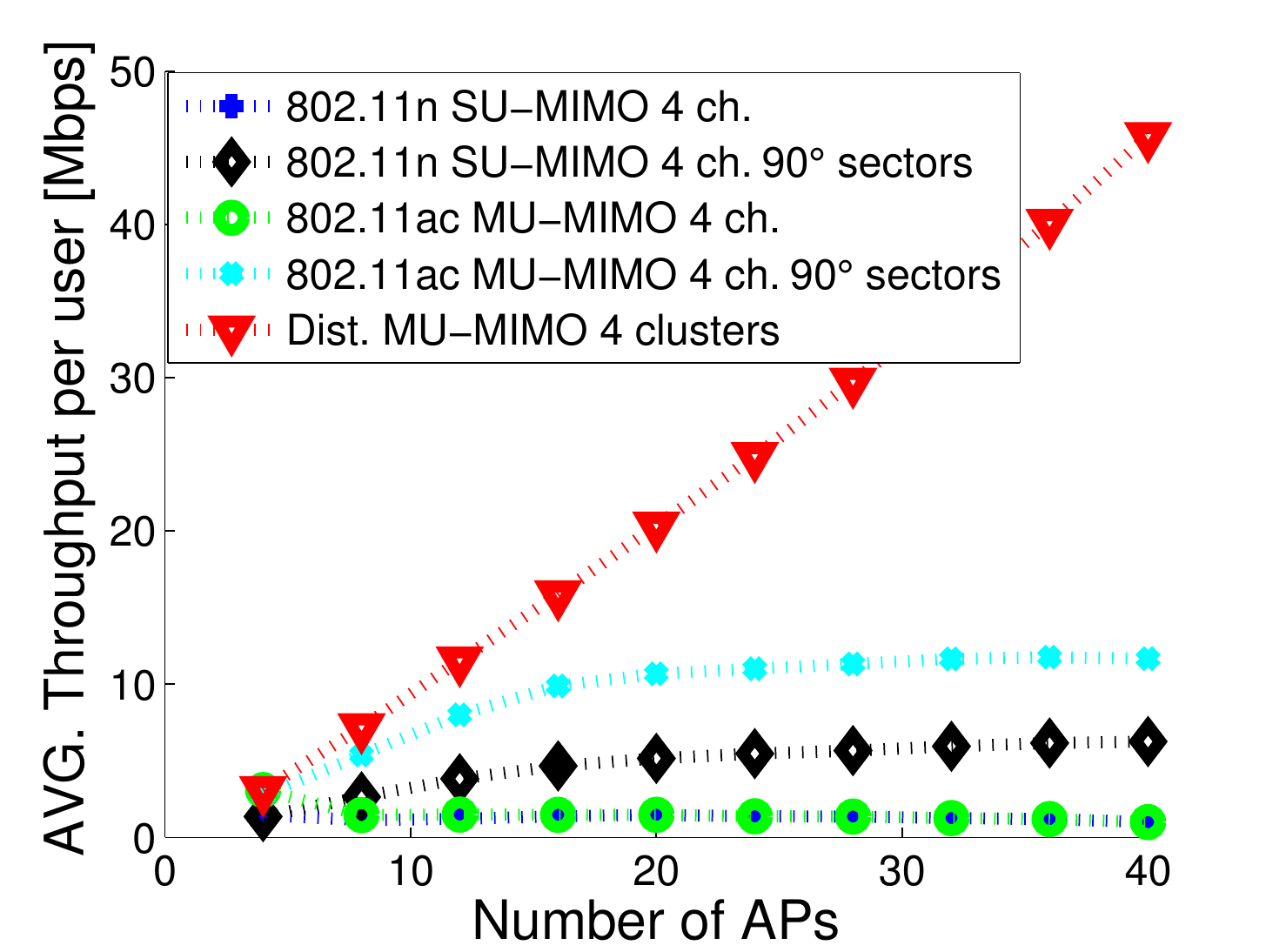}
   \captionof{figure}{Average throughput per user in a conference hall with and without sectorization with transmit power of 90dB and no CCA.}
   \label{fig:sectors}
\end{minipage}
\quad
\begin{minipage}[t]{.3\textwidth}
   \includegraphics[width=\textwidth]{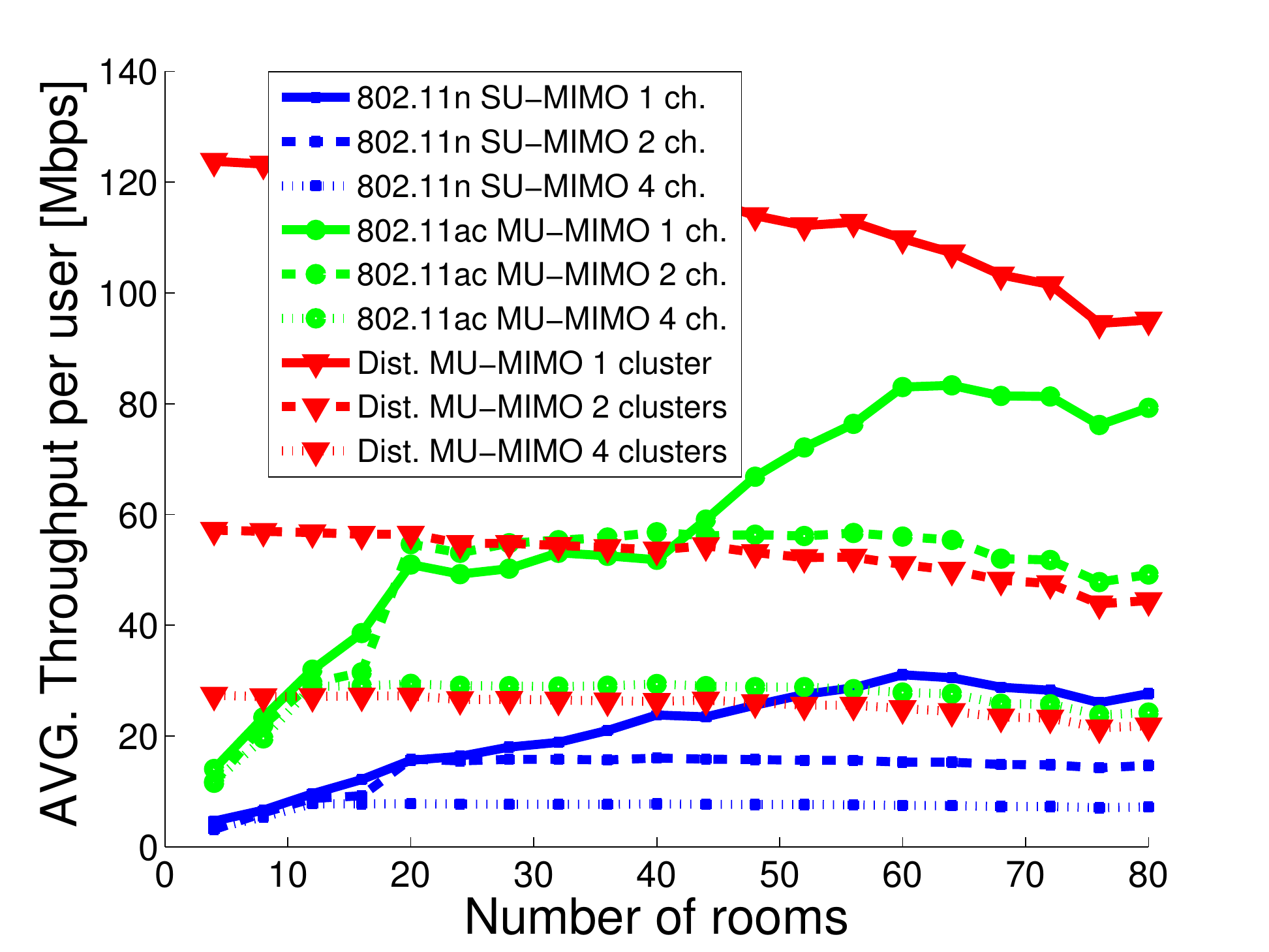} % requires the graphicx package
   \captionof{figure}{ Average throughput per user for an office building with rooms with transmit power of 90dB and $\text{CCA}=10$dB.}
   \label{fig:rooms1}
\end{minipage}
\quad
\begin{minipage}[t]{.3\textwidth}
   \includegraphics[width=\textwidth]{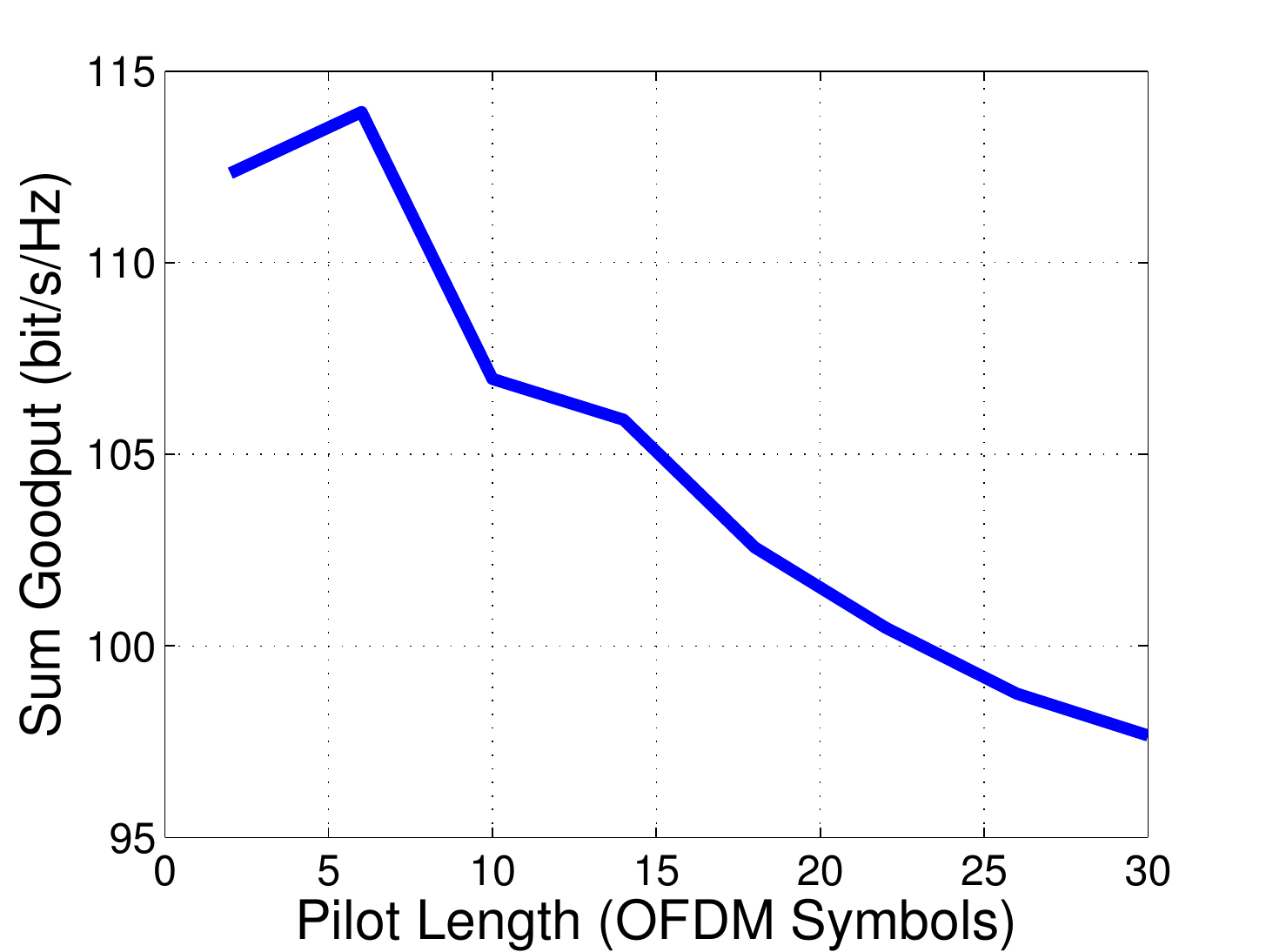}
   \captionof{figure}{Distributed MIMO overhead tradeoff due to synchronization and calibration requirements.}
   \label{fig:sync-overhead}
\end{minipage}
\end{figure*}

\begin{figure*}
   \centering
\begin{subfigure}[t]{.3\textwidth}
   \includegraphics[width=\textwidth]{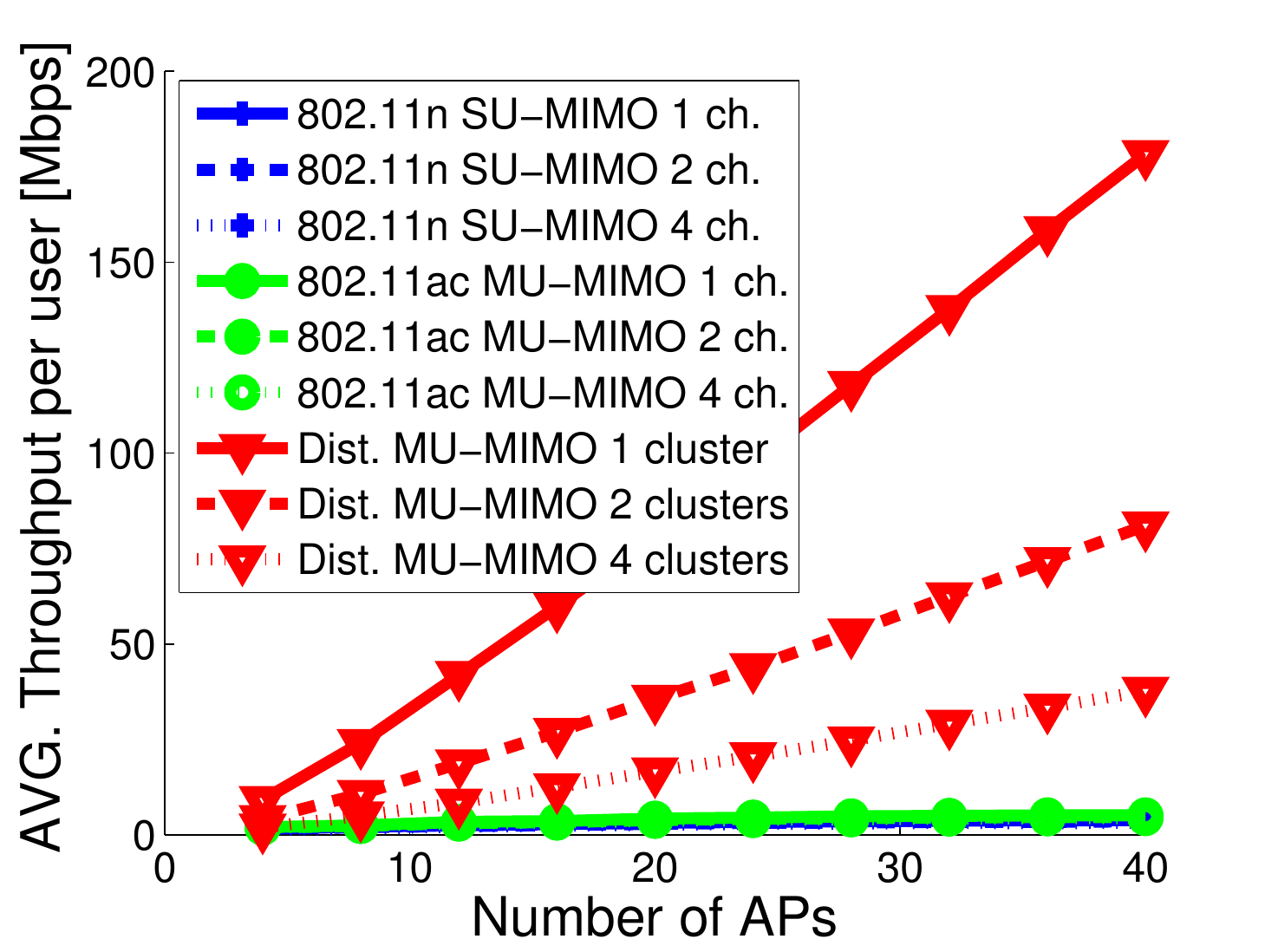} % requires the graphicx package
   \caption{ All technologies with $\text{CCA}=10$dB}
   \label{fig:open1}
\end{subfigure}
\quad
\begin{subfigure}[t]{.3\textwidth}
   \includegraphics[width=\textwidth]{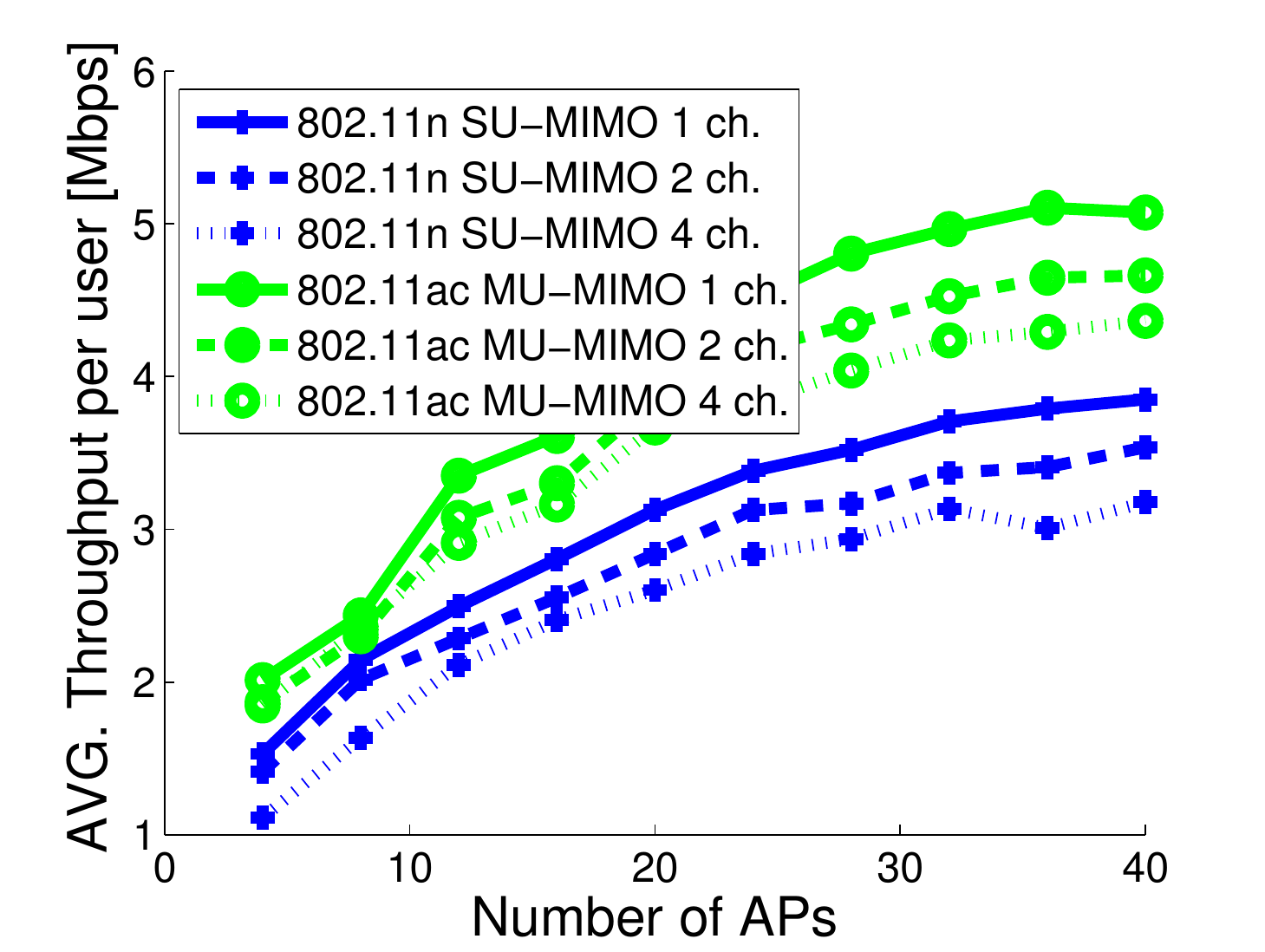} % requires the graphicx package
   \caption{ 802.11n SU-MISO and  802.11ac MU-MIMO with $\text{CCA}=10$dB}
   \label{fig:open2}
\end{subfigure}
\quad
\begin{subfigure}[t]{.3\textwidth}
   \includegraphics[width=\textwidth]{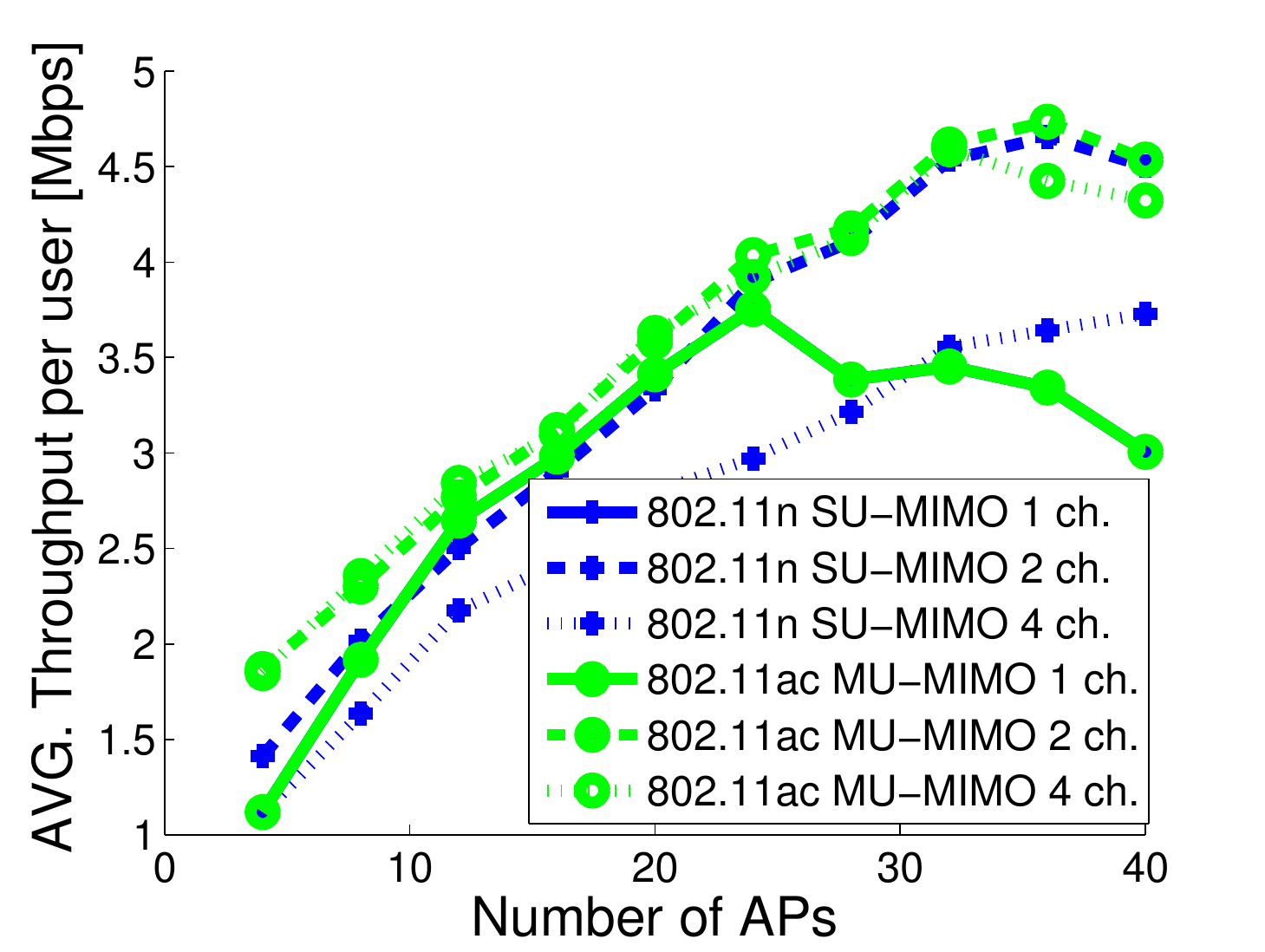} % requires the graphicx package
   \caption{  802.11n SU-MISO and  802.11ac MU-MIMO with no CCA}
   \label{fig:open3}
\end{subfigure}
\caption{Average throughput per user for an open-floor office plan with transmit power of 90dB.}
\end{figure*}

\begin{figure*}
   \centering
\begin{subfigure}[t]{.3\textwidth}
   \includegraphics[width=\textwidth]{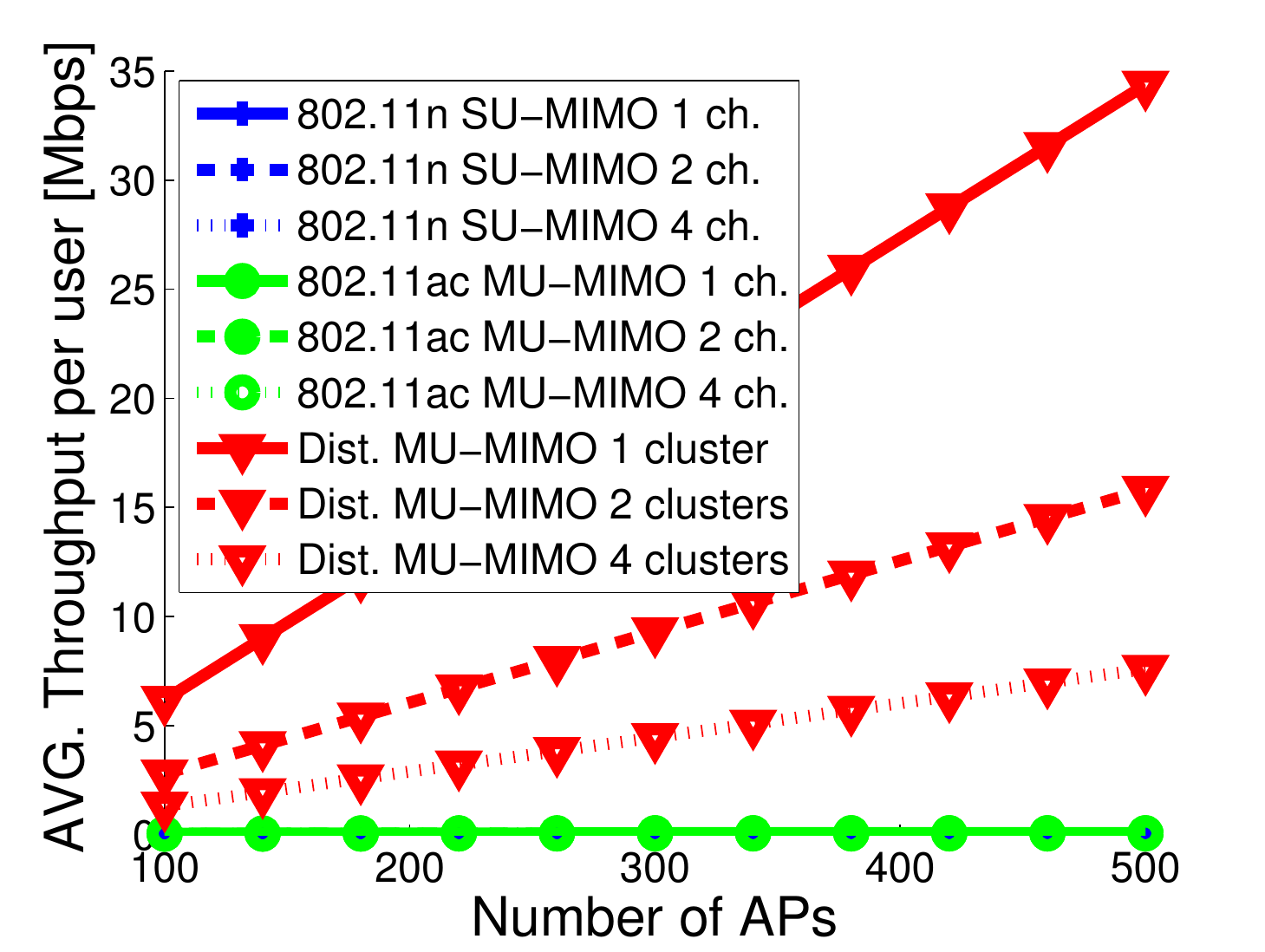} % requires the graphicx package
   \captionof{figure}{ All technologies with $\text{CCA}=2$dB.}
   \label{fig:stadium1}
\end{subfigure}
\quad
\begin{subfigure}[t]{.3\textwidth}
   \includegraphics[width=\textwidth]{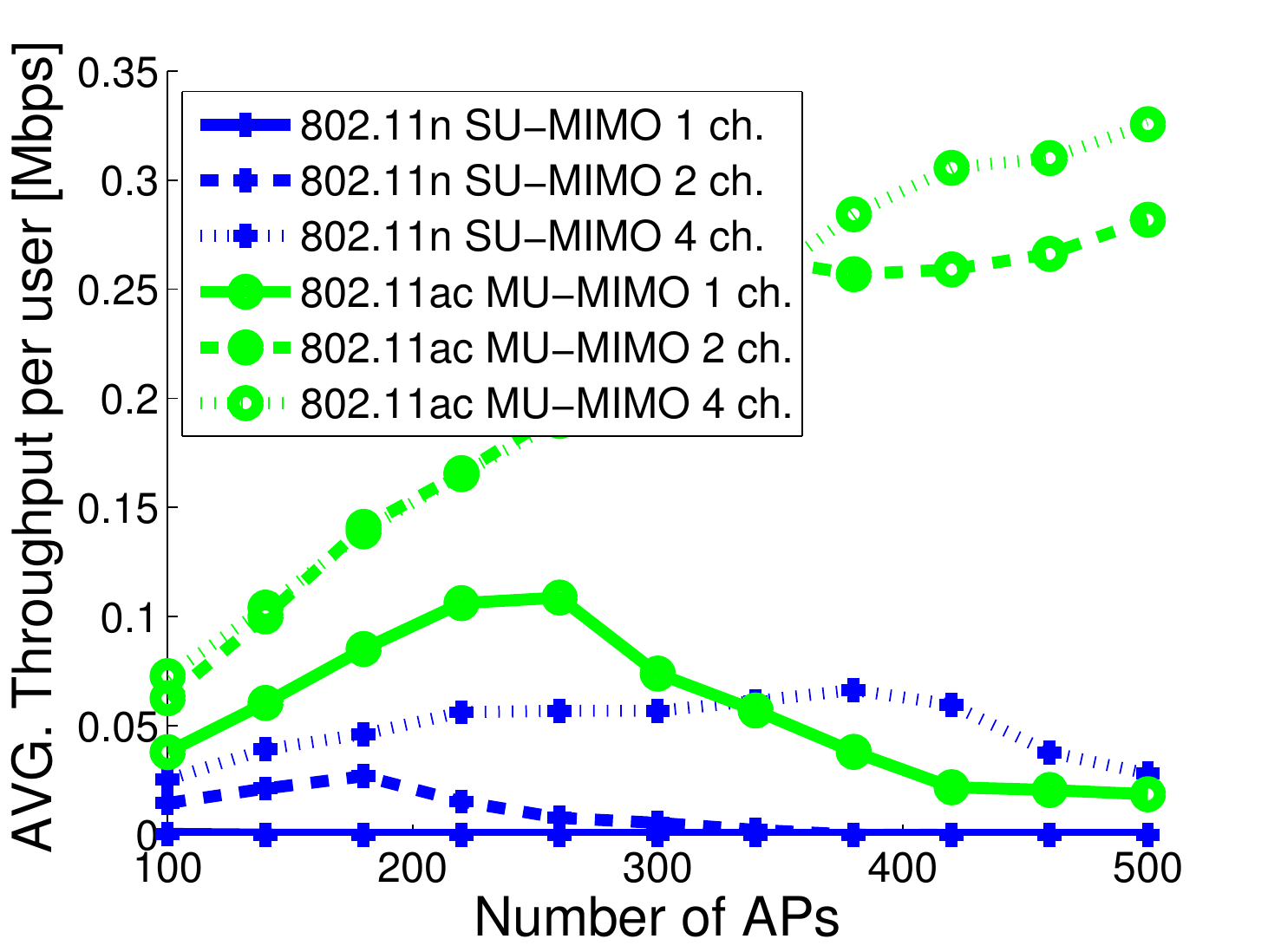} % requires the graphicx package
   \captionof{figure}{ 802.11n SU-MISO and 802.11ac MU-MIMO with no CCA.}
   \label{fig:stadium2}
\end{subfigure}
\quad
\begin{subfigure}[t]{.3\textwidth}
\includegraphics[width=\textwidth]{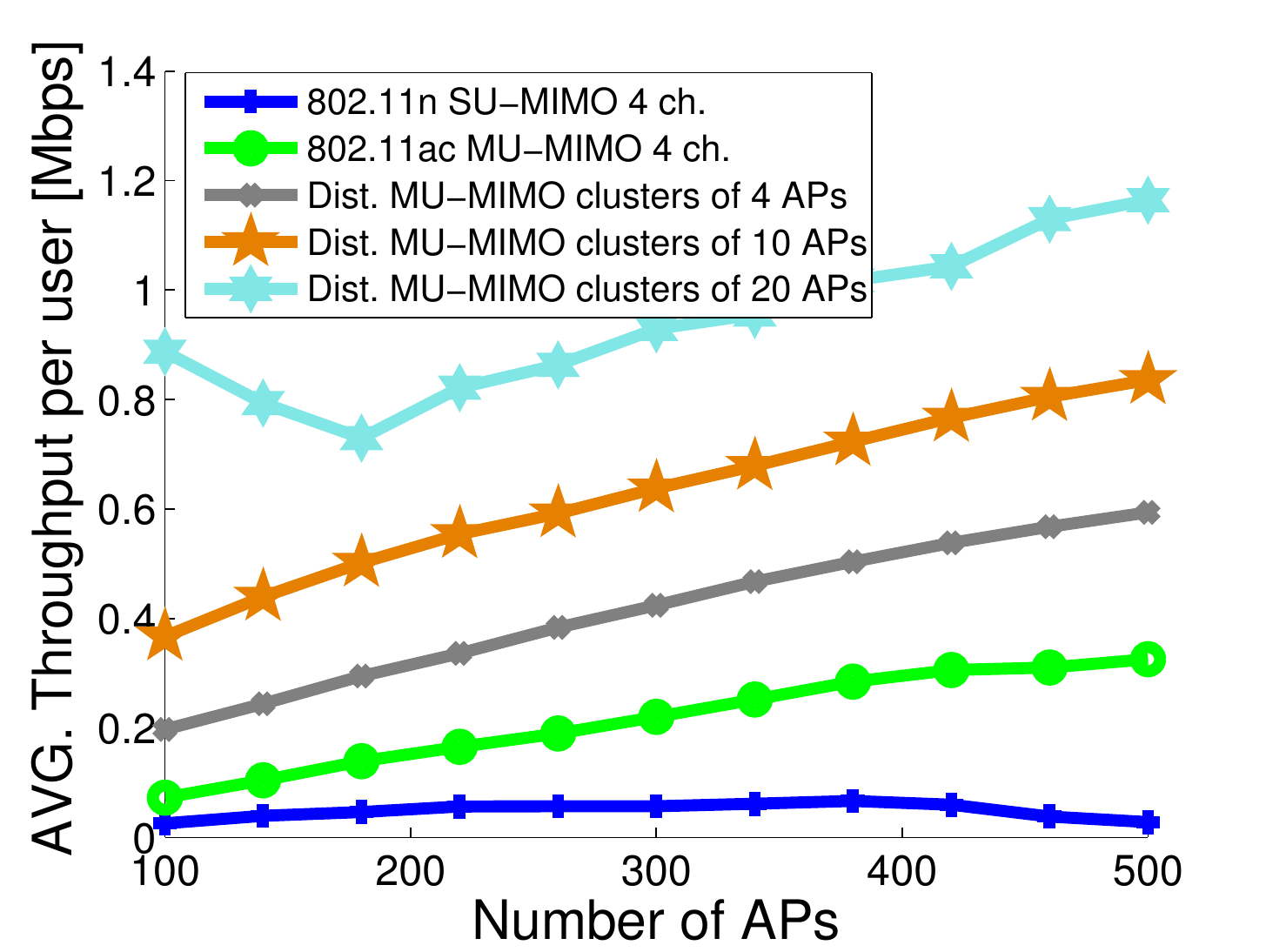} 
   \captionof{figure}{Average throughput per user for a stadium with small clusters consisting of neighboring APs only.}
   \label{fig:cluster}
\end{subfigure}
\caption{Average throughput per user for a stadium with $20000$ users and transmit power of 90dB.}
\end{figure*}

{\bf Sectorization gains:} In the previous results we saw that the distributed MU-MIMO technology significantly outperforms the presented non-coordinated schemes that quickly become 
interference limited as the number of APs grows. To mitigate the interference between APs that operate on the same channel we can  use a \textit{sectorized} deployment. Such 
deployments can commonly be found in cellular networks, where the antennas used have a sub-circular sector shaped radiation pattern \cite{cellular_sector}, and simplified variations 
are slowly making their appearance in enterprise WiFi deployments. 
We present here a simple scheme to illustrate the gains of mitigating the interference for non-coordinated technologies with such an approach. We shape the radiation patterns of all APs 
in the conference hall topology (Figure \ref{fig:amphi_topo}) such that they only radiate energy to a $90^o$ sector, oriented towards the same direction for all APs. This simple 
approach significantly improves the performance of SU-MIMO 802.11n and MU-MIMO 802.11ac technologies as can be seen in Figure \ref{fig:sectors}, by a factor of 3x and 7x 
respectively. Note that distributed MU-MIMO, which uses omnidirectional antennas, still performs significantly better, especially in a 1 cluster setup. (Only the 4 cluster setup is shown in 
this plot to better distinguish the performance of the other schemes).
%(CSMA was turned off for this comparison to present the gains of sectorization on it's own). 

\subsection{Open-floor office building}

The open-floor office scenario shares a few common characteristics with the conference hall but has some interesting differences. We define an office floor of dimensions $160\text{m}
\times 23\text{m}$ that is occupied by cubicles (essentially no walls obscure the line of sight from AP-to-AP or AP-to-user). APs are placed equally spaced in two rows and users are 
scattered uniformly and at random in the whole floor (see Figure \ref{fig:topo_openfloor}). The distributed MU-MIMO system clearly outperforms the non-coordinated solution in terms of 
average per user throughput as can be seen in Figure \ref{fig:open1}. It is interesting to notice that in contrast to the previous case, increasing the number of APs even up to 40 gives 
some gains for the non-coordinated systems under the CSMA scheduling scenario (see Figure \ref{fig:open2}). From Figure \ref{fig:open3} it is apparent that when the CCA threshold is 
removed all together, the regime becomes interference limited as before. This can be seen by the rate degradation of the single channel system when adding more than 20 APs in the 
topology. 

At this point we must emphasize the impact of using quantized rates in comparison to the information theoretic Gaussian rates. Although the latter offer a starting point to solve complex 
optimization problems and attain intuition 
for physical layer effects, they might lead to misleading intuition if not combined with an actual quantized mapping. Specifically, the scenarios with no CCA threshold (see Figures 
\ref{fig:open3} for the open-floor office building 
and \ref{fig:amphi3} for the conference hall scenario) would look significantly different in their respective Gaussian versions. The rates would still saturate, but even at very low SINR, 
when the MCS mapping would give rates 
equal to zero, the Gaussian rate would assign some positive rate to every user. Further, increasing the number of APs increases the number of users served simultaneously (up to some 
limit), so the use of Gaussian rates 
leads to computing a higher average rate per user than what a quantized rate scheme would give.

%The interesting difference with the previous scenario is that the gap between 802.11ac MU-MIMO and 802.11n SU-MISO has been significantly decreased even with the case of a 
%CCA threshold of $10$dB as can be seen in Figure \ref{fig:open1}. Indeed, although the interference is lower

\subsection{Office building floor with rooms}
\label{sec:office_rooms}

A typical office building floor with two rows of rooms and a corridor in between them (see Figure \ref{fig:topo_rooms}) is analyzed with our model. The length of the floor is 160m, the 
width of each room is 10m, and the corridor width is 3m. For this series of experiments we fix the number of APs to $20$ and users to $200$ and investigate the behavior of the network 
as the number of rooms is varied. APs are placed in a canonical way to cover the whole area and users are scattered uniformly at random in the whole floor.

The increase of the number of rooms, while the floor dimensions and number of APs remains the same, essentially adds additional barriers (walls) between APs. This, in return, 
effectively increases the pathloss between APs that share the same channel and thus significantly lowers the interference. On the other hand, the users fall out of line-of-sight with more 
APs, and in fact their pathlosses to many of the APs substantially increase due to the walls in between them. Thus, the multiplexing gains of distributed MU-MIMO disappear shortly after 
the number of rooms increases.

In Figure \ref{fig:rooms1} we can see that as the APs become better separated by adding walls between them, the gains of distributed MU-MIMO are negligible and eventually trail 
behind the localized MU-MIMO scheme. This is expected since in this scenario, only the gain with the AP assigned to each user is significant and thus, trying to beam form to users from 
distant APs, that are not in their transmit range, actually hurts the system instead of providing multiplexing gains. 
%This phenomenon has been verified by our simulations in Section \ref{section:validation} (see Figures \ref{fig:validation5} and \ref{fig:validation6}).

% has been witnessed before in simulations \cite{globecom_zander} and 

\subsection{Stadium}

Having the analytic model makes it easy and fast to compute the expected average user throughput in even larger scenarios as would be in the case of a football stadium (with a radius 
of 100m) during a big game (see Figure 
\ref{fig:stadium_topo} for topology and channel assignment). We can imagine that at moments of high interest, for instance during a questionable refereeing decision or a touchdown, a 
large percentage of the in stadium fans 
might want to see a replay or a different camera feed online. In Figure \ref{fig:stadium1} we can see the average throughput per user for a stadium of dimensions $200\text{m} \times 
200\text{m}$, with 20000 users trying to 
access the internet and a range of APs. It is obvious in this case that only distributed MU-MIMO has the potential to provide an acceptable user service under such a challenging yet 
typical scenario. Moreover, if the CCA 
threshold is completely removed we again end up in an interference limited scenario where non-coordinated technologies suffer from rate saturation/degradation as the number of APs 
increases (see Figure \ref{fig:stadium2}).

In such a large deployment, managing to coordinate hundreds of APs at the level of accuracy the distributed MU-MIMO architecture requires might be very challenging if not impossible 
in practice. Thus, a different approach consists of clustering only a small number of closely located APs in a distributed MU-MIMO cluster, and operating a large number of such clusters 
independently to cover the whole area. This has the effect of having clusters operating in the same frequency and not cooperating, thus creating interference to each other. Such 
deployments can also be efficiently analyzed using our model and the results can be seen in Figure \ref{fig:cluster}. We see that the performance of distributed MU-MIMO when APs are 
grouped in small clusters consisting of neighboring APs is significantly lower than that of a system coordinating all hundreds of APs together (compare Figures  \ref{fig:stadium1} and 
\ref{fig:cluster}). That said, it is still a lot better than uncoordinated MU-MIMO. 
%(CSMA was turned off for this comparison to present the gains of clusterization on it's own).

\subsection{ Signaling overhead}

%synchronization and calibration
\label{sec:appendix}
{\bf CSIT overhead:} It is worth noting that all the PHY schemes considered require the knowledge of channel state information at the transmitter side (CSIT) \cite{molisch-book}. 
The overhead of the collection of CSIT exists and is similar in all aforementioned systems, but differs per topology and thus can only be taken into account in a per-scenario basis, 
it is thus not included in the analytic model. It is however straightforward to compute and discount for such an overhead based on the 802.11n/ac standards \cite{802.11ac} and on 
some recent work of ours \cite{ryan-twc}.

{\bf Distributed MU-MIMO overhead:} Distributed MU-MIMO requires additional overhead to coordinate the APs in the cluster and allow them to operate as a single, virtual AP. 
In particular, timing and frequency synchronization (necessary for coherent distributed ZFBF) 
and uplink-downlink reciprocity (necessary to learn the downlink channel matrix from the users uplink data packets) 
must be implemented through an efficient and scalable synchronization and calibration protocol. These problems 
have been recently addressed in depth in our work \cite{ryan-twc}, as well as in a number of publications reporting 
software-defined radio prototype implementations of distributed MU-MIMO by us \cite{airsync_ton} and others \cite{rahul_jmb, nemox}.

Practical synchronization between the APs is imperfect and residual carrier frequency offsets (CFOs) lead to a rate degradation that gets progressively worse over the course of 
transmission. %These residual CFOs depend on the specific frequency offset estimation scheme and on the SNR of the channels between the APs \cite{ryan-twc}.
Developing an analytical characterization of the tradeoff between the achievable rates and the synchronization overhead is difficult, since it is highly dependent upon the selection 
of ``anchor'' or leader nodes, and it involves non-convex optimization \cite{ryan-twc}. As such, this problem can be solved on a per-topology basis using simulations. %such as 
those that have been developed as part of our prior work \cite{ryan-twc}.

%Practical synchronization between the APs is imperfect, and the residual carrier frequency offsets (CFOs) between the APs lead to a rate degradation that becomes progressively 
%worse over the course of transmission, because the CFOs introduce a phase offset between the transmitting APs, which grows with the duration of the downlink data packet. These 
%residual CFOs  depend on the specific frequency offset estimation scheme and on the SNR of the channels between the APs \cite{ryan-twc}. Since the SNR between APs is 
%determined by fixed by the network topology, here we focus on the trade-off between the synchronization pilot sequence length and the achievable rates. Developing an analytical 
%characterization of the tradeoff between the achievable rates and the synchronization overhead is difficult, since it is highly dependent upon the selection of ``anchor'' or leader 
%nodes, and it involves on-convex optimization \cite{ryan-twc}. As such, this problem is solved on a per-topology basis via simulation, for the purposes of this paper.  

Figure \ref{fig:sync-overhead} illustrates the tradeoff between achievable rates and pilot overhead.  
This simulation is for the $N_a = 20$ APs and  $K = 200$ users conference hall scenario (see Section \ref{sec:conf_hall}), 
where clusters of $B = 5$ APs form distributed MU-MIMO systems (see Section \ref{dmumimo} for details of clustering).  
We see that the optimal length of the synchronization sequence is 6 OFDM symbols, 
after which the longer sequences give diminishing returns in terms of synchronization quality, 
while consuming more overhead in the synchronization protocol and therefore reducing the system goodput. 
Thus each cluster needs 30 OFDM symbols of overhead for synchronization in the worst case, because the synchronization pilot signals need to be orthogonal in time.
%We note that the overhead can be further reduced by having only some APs transmit synchronization sequences (the anchor APs), and having the others just synchronizing in a 
%master-slave way. 
%The accurate optimization tradeoffs between the number of anchors, the sequence length, and the achievable rates, is topology-dependent and it is outside the scope of this 
%work.

In addition to synchronization overhead, distributed MIMO utilizes calibration between the APs in order to achieve 
uplink-downlink reciprocity \cite{ryan-twc, argos}.  
For a distributed MU-MIMO system, due to the residual CFOs, calibration pilot sequences must also be repeated at every downlink data slot \cite{ryan-twc}. %(unlike the case of 
%\cite{argos} where all antennas are driven by the same clock). 
%As with synchronization, calibration is achieved by having the APs exchange pilot signals in the form of OFDM symbols. 
In the previous example, we have determined that the optimal calibration sequence length is 4 OFDM symbols, for a total of 20 OFDM symbols per cluster.  
In general, for a given network configuration and distributed MU-MIMO clustering, it is possible to optimize ``off-line'' the overhead incurred by synchronization and calibration, and 
apply this overhead as a discount factor to 
calculate the goodput of the system.

%\section{Channel and User-AP Assignment} 
%\label{sec:appendix2}
%APs are ordered at random and choose their channel sequentially, such that AP $\pi(i)$ makes its choice at step $i$, 
%where $\pi$ denotes the random ordering permutation, 
%and chooses channel $c_{\pi(i)} \in \Cc$ where $ \Cc$ is the collection of non-overlapping channels used with an arbitrary numbering from such that:
%$$ c_{\pi(i)} = \arg \min _{c \in \Cc} \left \{ \sum_{j < i : \pi(j) \in {\Ac_{c}}} I_{\pi(i),\pi(j)}^c \right \} $$
%where $I_{\ell,r}^c$ denotes the interference caused on AP $\ell$ by AP $r$ in channel $c$, 
%and  $\Ac_{c}$ denotes the set of APs already assigned to channel $c$. 

%
%\begin{figure}
%   \centering
%   \includegraphics[width=.4\textwidth]{stadium_clust} % requires the graphicx package
%   \caption{ Average throughput per user for a stadium with clustered distributed MU-MIMO}
%   \label{fig:cluster}
%\end{figure}
%

\section{Conclusions}

In this paper we have introduced an accurate and practical analytical model for next generation wireless networks. We have applied the model in a variety
of real-world scenarios and investigated the effect of all important system/design parameters to performance. 
An important result from this study is the significant performance gains that distributed MU-MIMO has over non coordinated approaches 
like concentrated MU-MIMO. That said, distributed MU-MIMO requires not only CSIT, but also tight synchronization and thus it incurs additional overhead.
Also, in the presence of walls and other barriers which reduce inter-cell interference its gains are small, and, when considering a cap on the total
number of APs that can be efficiently coordinated in practice, again its gains reduce. Other interesting outcomes of this work are the sizable gains 
from sectorization, though to achieve those gains in practice, expensive front ends would have to be used (such front ends are typical in the context of
cellular base stations but make less sense in the context of WiFi APs), as well as the speed by which non coordinated approaches become 
interference-limited resulting in no additional gains as more APs are added on the network.
As a final point, it is evident from the depth and breath of the analysis of various practical scenarios in this work that our analytical model can
be used to guide the deployment of future wireless networks and to optimize their performance. 

%\FloatBarrier

\bibliographystyle{IEEEtran}
\bibliography{IEEEabrv,wlan_comp_abbreviated}

%\input{bios}
%\appendices
%\input{overhead}

\end{document}